\documentclass[longauth]{aa} % for the long lists of affiliations 
\usepackage{graphicx}
\usepackage{txfonts}
\usepackage[normalem]{ulem}

\begin{document}

   \title{NGTS-30\,b/TOI-4862\,b: An $\sim$1 Gyr old 98-day transiting warm Jupiter\footnote{}}

   \titlerunning{A 98-day warm Jupiter}

   \authorrunning{Battley et al.}

   %\subtitle{I. Overviewing the $\kappa$-mechanism}

   \author{M.P. Battley\inst{1},
          K.A. Collins\inst{2},
          S. Ulmer-Moll\inst{1},\inst{3},
          S.N. Quinn\inst{2},
          M. Lendl\inst{1},
          S. Gill\inst{4,}\inst{5},
          R. Brahm\inst{6,}\inst{7,}\inst{8},
          M.J.~Hobson\inst{1},
          H.P.~Osborn\inst{3},
          A.~Deline\inst{1},
          J.P.~Faria\inst{1},
          A.B.~Claringbold\inst{3,}\inst{4},
          H. Chakraborty\inst{1},
          K.G.~Stassun\inst{9},
          C. Hellier\inst{10},
          D. R. Alves\inst{11},
          C. Ziegler\inst{12},
          D.R. Anderson\inst{13},
          I.~Apergis\inst{4,}\inst{5},
          D.J.~Armstrong\inst{4,}\inst{5},
          D. Bayliss\inst{4,}\inst{5},
          Y.~Beletsky\inst{14},
          A. Bieryla\inst{2},
          F. Bouchy\inst{1},
          M.R.~Burleigh\inst{15},
          R.P. Butler\inst{16},
          S.L.~Casewell\inst{15},
          J.L.~Christiansen\inst{17},
          J.D.~Crane\inst{18},
          P.A. Dalba\inst{19},
          T.~Daylan\inst{20},
          P.~Figueira\inst{1},
          E.~Gillen\inst{21},
          M.R.~Goad\inst{15},
          M.N.~Günther\inst{22},
          B.A.~Henderson\inst{15},
          T.~Henning\inst{23}
          J.S.~Jenkins\inst{24,}\inst{25},
          A.~Jordán\inst{6,}\inst{7,}\inst{8},
          S.~Kanodia\inst{16},
          A.~Kendall\inst{15},
          M.~Kunimoto\inst{26,}\inst{27},
          D.W.~Latham\inst{2},
          A.M. Levine\inst{26,},
          J.~McCormac\inst{4,}\inst{5},
          M.~Moyano\inst{13},
          A.~Osborn\inst{4,}\inst{5},
          D.~Osip\inst{14},
          T.A. Pritchard\inst{28},
          A. Psaridi\inst{1},
          M. Rice\inst{29},
          J.E.~Rodriguez\inst{30},
          S. Saha\inst{24,}\inst{25},
          S.~Seager\inst{26,}\inst{31,}\inst{32},
          S.A.~Shectman\inst{18},
          A.M.S.~Smith\inst{33},
          J.K.~Teske\inst{34},
          E.B. Ting\inst{35},
          S.~Udry\inst{1},
          J.I.~Vines\inst{13},
          C.A.~Watson\inst{36},
          R.G.~West\inst{4,}\inst{5},
          P.J.~Wheatley\inst{4,}\inst{5},
          J.N. Winn\inst{37},
          S.W.~Yee\inst{37}
          \and
          Y. Zhao\inst{1}
          }

   \institute{Observatoire Astronomique de l’Université de Genève, Chemin Pegasi 51, CH-1290 Versoix, Switzerland
        \and    
            Center for Astrophysics | Harvard \& Smithsonian, 60 Garden Street, Cambridge, MA 02138, USA
        \and
            Physikalisches Institut, University of Bern, Gesellsschaftstrasse 6, 3012 Bern, Switzerland
        \and
            Department of Physics, University of Warwick, Gibbet Hill Road, Coventry CV4 7AL, UK
        \and
            Centre for Exoplanets and Habitability, University of Warwick, Gibbet Hill Road, Coventry CV4 7AL, UK
        \and
            Facultad de Ingeniería y Ciencias, Universidad Adolfo Ibáñez, Av. Diagonal las Torres 2640, Peñalolén, Santiago, Chile
        \and
            Millennium Institute for Astrophysics, Chile
        \and
            Data Observatory Foundation, Chile
        \and
            Department of Physics and Astronomy, Vanderbilt University, Nashville, TN 37235, USA
        \and
            Astrophysics Group, Keele University, Staffordshire, ST5 5BG, UK
        \and
            Departamento de Astronom\'{i}a, Universidad de Chile, Casilla 36-D, Santiago, Chile
        \and
            Department of Physics, Engineering and Astronomy, Stephen F. Austin State University, 1936 North St, Nacogdoches, TX 75962, USA
        \and
            Instituto de Astronom\'ia, Universidad Cat\'olica del Norte, Angamos 0610, 1270709, Antofagasta, Chile
        \and
            Las Campanas Observatory, Carnegie Institution for Science, Colina el Pino, Casilla 601 La Serena, Chile
        \and
            School of Physics and Astronomy, University of Leicester, Leicester LE1 7RH,  UK
        \and
            Earth and Planets Laboratory, Carnegie Institution for Science, 5241 Broad Branch Road, NW, Washington, DC 20015, USA
        \and
            Caltech/IPAC-NASA Exoplanet Science Institute, 770 S. Wilson Avenue, Pasadena, CA 91106, USA
        \and
            The Observatories of the Carnegie Institution for Science, 813 Santa Barbara St., Pasadena, CA 91101, USA
        \and
            Department of Astronomy and Astrophysics, University of California, Santa Cruz, CA 95064, USA
        \and
            Department of Physics and McDonnell Center for the Space Sciences, Washington University, St. Louis, MO 63130, USA
        \and
            Astronomy Unit, Queen Mary University of London, Mile End Road, London E1 4NS, UK
        \and
            European Space Agency (ESA), European Space Research and Technology Centre (ESTEC), Keplerlaan 1, 2201 AZ Noordwijk, The Netherlands
        \and
            Max-Planck-Institut für Astronomie, Königstuhl 17, 69117 Heidelberg, Germany
        \and
            Instituto de Estudios Astrof\'isicos, Facultad de Ingenier\'ia y Ciencias, Universidad Diego Portales, Av. Ej\'ercito 441, Santiago, Chile
        \and
            Centro de Astrof\'isica y Tecnolog\'ias Afines (CATA), Casilla 36-D, Santiago, Chile
        \and
            Department of Physics and Kavli Institute for Astrophysics and Space Research, Massachusetts Institute of Technology, 77 Massachusetts Ave, Cambridge, MA 02139, USA
        \and
            Juan Carlos Torres Fellow
        \and
            NASA Goddard Space Flight Center, 8800 Greenbelt Road, Greenbelt, MD 20771, USA
        \and
            Department of Astronomy, Yale University, New Haven, CT 06511, USA
        \and
            Center for Data Intensive and Time Domain Astronomy, Department of Physics and Astronomy, Michigan State University, East Lansing, MI 48824, USA
        \and
            Department of Earth, Atmospheric and Planetary Sciences, Massachusetts Institute of Technology, Cambridge, MA 02139, USA
        \and
            Department of Aeronautics and Astronautics, MIT, 77 Massachusetts Avenue, Cambridge, MA 02139, USA
        \and
            Insitute of Planetary Research, German Aerospace Center (DLR), Rutherfordstr. 2, 12489 Berlin, Germany
        \and
            Earth and Planets Laboratory, Carnegie Institution for Science, 5241 Broad Branch Rd NW, Washington, DC 20015
        \and
            NASA Ames Research Center, Moffett Field, CA 94035 USA
        \and
            Astrophysics Research Centre, Queen’s University Belfast, Belfast BT7 1NN, UK
        \and
            Department of Astrophysical Sciences, Princeton University, Princeton, NJ 08544, USA
            }
            
   \date{Received Jan 2024; accepted Mar 2024}

% \abstract{}{}{}{}{} 
% 5 {} token are mandatory
 
  \abstract
    {Long-period transiting exoplanets bridge the gap between the bulk of transit- and Doppler-based exoplanet discoveries, providing key insights into the formation and evolution of planetary systems. The wider separation between these planets and their host stars results in the exoplanets typically experiencing less radiation from their host stars; hence, they should maintain more of their original atmospheres, which can be probed during transit via transmission spectroscopy. Although the known population of long-period transiting exoplanets is relatively sparse, surveys performed by the Transiting Exoplanet Survey Satellite (TESS) and the Next Generation Transit Survey (NGTS) are now discovering new exoplanets to fill in this crucial region of the exoplanetary parameter space.}
   {This study aims to characterise a new long-period transiting exoplanet by following up on a single-transit candidate found in the TESS mission.}
   {The TOI-4862 system was monitored using a combination of photometric instruments (TESS, NGTS, and EulerCam) and spectroscopic instruments (CORALIE, FEROS, HARPS, and PFS) in order to determine the period, radius, and mass of the long-period transiting exoplanet NGTS-30\,b/TOI-4862\,b. These observations were then fitted simultaneously to determine precise values for the properties and orbital parameters of the exoplanet, as well as the refined stellar parameters of the host star.}
   {We present the discovery of a long-period (P = 98.29838$\pm$0.00010 day) Jupiter-sized (0.928$\pm$0.032 R$_\mathrm{J}$; 0.960$\pm$0.056 M$_\mathrm{J}$) planet transiting a 1.1 Gyr old G-type star, one of the youngest warm Jupiters discovered to date. NGTS-30\,b/TOI-4862\,b has a moderate eccentricity (0.294$^{+0.014}_{-0.010}$), meaning that its equilibrium temperature can be expected to vary from 274$^{+30}_{-46}$ K to 500$^{+55}_{-84}$ K over the course of its orbit. Through interior modelling, NGTS-30\,b/TOI-4862\,b was found to have a heavy element mass fraction of $\rm 0.23^{+0.05}_{-0.06}$ and a heavy element enrichment ($\rm Z_{p} / Z_{\star} $) of $20^{+5} _{-6}$, making it metal-enriched compared to its host star. }
   {NGTS-30\,b/TOI-4862\,b is one of the youngest well-characterised long-period exoplanets found to date and will therefore be important in the quest to understanding the formation and evolution of exoplanets across the full range of orbital separations and ages.}


   \maketitle
\renewcommand*{\thefootnote}{\fnsymbol{footnote}}
\footnotetext{* This paper includes data gathered with the 6.5 metre Magellan Telescopes located at Las Campanas Observatory, Chile.}
\renewcommand*{\thefootnote}{\arabic{footnote}}
%
%-------------------------------------------------------------------

\section{Introduction}

% Why long period exoplanets
Three decades of exoplanet discovery have revealed a population of exoplanets far more diverse than originally anticipated. This diversity is shaped by a combination of formation processes \citep[e.g.][]{Boss1997GiantInstability,Safronov1972EvolutionPlanets.,Lissauer1993PlanetFormation}, planetary migration \citep[e.g.][]{Ward1986DensityTorque,Lin1986OnProtoplanets,Kozai1962SecularEccentricity,Lidov1962TheBodies}, atmospheric evolution \citep[e.g.][]{Owen2019AtmosphericExoplanets}, and dynamical interactions \citep[e.g.][]{Ida2010TowardStars}, which leave distinct imprints on the overall distribution of exoplanets. One particularly interesting type of exoplanet is `warm Jupiters', planets with masses and radii approximately that of Jupiter and periods of 10-200 days, which bridge the gap between the short-period `hot Jupiters' (P< 10 days) commonly found in transit surveys and the gas giants in the Solar System. These exoplanets are particularly interesting for studying the evolution and atmospheres of giant planets, as, like the giants in the Solar System, they are found much farther from their host stars and hence are not subject to the intense radiation environments of hot Jupiters \citep{Showman2020AtmosphericDwarfs}. Likewise, this increased distance means that they are tidally detached from their host stars, so they can offer more clues about their formation environments \citep{Rice:2022}. This aspect is especially important to explore as there is some evidence that they may have undergone less migration than close-in exoplanets \citep{Boley2016THEPERIODS,Wu2011SECULARJUPITERS,Bitsch2019FormationGiants}. %This class of planets also covers a distinct gap in the current population of exoplanets above $\sim$0.2$M_\mathrm{Jup}$ and between $\sim$10-100 days, where the occurrence rate of large exoplanets appears to drop off dramatically \citep{Udry2003StatisticalExoplanets,Wittenmyer2010THEVALLEY}. 

% Eccentricity and age
The dominant drivers behind the formation and evolution of warm Jupiters are still not well understood \citep[e.g.][]{Muller2023WarmOutlook}. Three particularly important parameters that can be used to constrain these processes are the density, eccentricity, and age of warm Jupiters, which directly probe both the current state, ongoing mechanisms, and timescale of this evolution. The precise density retrieved from measurements of the mass and radius of an exoplanet allows the bulk metallicity of these planets to be estimated and, through comparison to synthetic models, allows a first look at the current internal structure of the exoplanet \citep[e.g.][]{Mordasini2012CharacterizationFormation,Mordasini2015GlobalEvolution}. Meanwhile, eccentricity measurements give insight into how warm Jupiters reached their current orbital position, with low-eccentricity exoplanets best explained by in situ formation or disk migration \citep[e.g.][]{Safronov1972EvolutionPlanets.,Lissauer1993PlanetFormation,Ward1986DensityTorque} and high-eccentricity exoplanets suggestive of high-eccentricity evolution mechanisms \citep[e.g.][]{Kozai1962SecularEccentricity,Lidov1962TheBodies}. Finally, measurements of the ages of such exoplanets allow them to be placed more effectively on their evolution tracks; this helps us understand whether the observed systems probe the final states of such planets or one that is still actively evolving. This is particularly crucial for high-eccentricity warm Jupiters, which may be in an intermediary evolution state on the way to hot Jupiter formation \citep{Fabrycky2007ShrinkingFriction}. To date, ages have been determined with precision better than 50\% for 24 well-constrained warm Jupiters. However, the majority of these are older (>5Gyr) and have periods of <60 days. To determine the dominant drivers shaping warm Jupiters, it is vital that a well-characterised sample of such exoplanets across their full evolutionary history be assembled.

%Combining different methods
Due to the geometric probability of a transit being inversely proportional to the orbital distance, the largest number of longer-period exoplanets have been found via the radial velocity (RV) method \citep[e.g.][]{Udry2003StatisticalExoplanets,Marmier2013ThePlanets,Borgniet2019ExtrasolarStars}. However, as different discovery methods allow for the determination of different planetary parameters, in-depth characterisation of exoplanets relies on combining different discovery methods. Crucially, the RV method alone cannot give the radius of an observed exoplanet and can only give the minimum mass of an exoplanet ($M_P \sin i$). The optimal way to solve this problem is to observe the transit of such a planet around its host star and determine the radius of the planet from the depth of the transit \citep{Borucki1984TheSystems}. Combining these methods provides us with  precise constraints on the density and composition of such planets. However, with fewer than 50 well-constrained warm Jupiters with precise masses and radii ($\sigma_{M_P} / M_P \leq$ 25\%; $\sigma_{R_P} / R_P \leq$ 8\%) currently known,\footnote{\url{https://exoplanetarchive.ipac.caltech.edu/}; accessed 17 November 2023} a larger sample of such planets is needed.

Long-period transiting exoplanets are particularly valuable because their atmospheres can be studied through transmission spectroscopy \citep{Seager2000TheoreticalTransits}, which directly probes their chemical composition. The moderate-temperature atmospheres of long-period planets (between well-studied hot Jupiters and Solar System gas giants) are predicted to host a range of observationally unconstrained atmospheric processes. Transmission spectroscopy of temperate gas giants can probe transitions in nitrogen chemistry \citep[e.g.][]{Fortney2020BeyondPlanets,hu2021photochemistry,Ohno2023NitrogenSpectra} and aerosol properties \citep[e.g.][]{Pont2008DetectionHubbleSpaceTelescope,gao2017sulfur,Brande2023CloudsExoplanets}, which are essential for understanding the entirety of the gas giant parameter space. By measuring the atmospheric composition of gas giants with different orbital architectures, we can hope to understand the formation and migration history of giant planets \citep[e.g.][]{Mordasini2012CharacterizationFormation,Ohno2023NitrogenSpectra}.

% Mono/duo chat
Unfortunately, directly looking for transits of known long-period planets discovered from RV measurements is difficult due to the decreased transit probability at these wider separations. A more efficient method is to look for long-period transiting exoplanet candidates with ongoing all-sky transit surveys such as the Transiting Exoplanet Survey Satellite (TESS; \citealt{Rickeretal.2014TheSatellite}) and measure their masses via RV follow-up. As TESS typically observes each region of the sky for only a month at a time, any long-period exoplanets are usually found as single-transit candidates -- or `mono-transits' -- requiring additional follow-up to determine their periods. \citet{Cooke2018SingleDetections} and \citet[using similar techniques as \citealt{Osborn2016SingleEstimation}]{Villanueva2019AnSatellite} each demonstrated that hundreds of such long-period planets should be detectable as single transits in the TESS Primary mission alone. 

These predictions have led to several mono-transit-specific searches in TESS data \citep[e.g.][]{Gill2020NGTS-11Event,Montalto2020AHemisphere} as well as those highlighted as TESS objects of interest \citep[TOIs;][]{Guerrero2021TheMission} in the main search completed by the TESS mission\citep{Rickeretal.2014TheSatellite,Jenkins2016TheCenter}. In order to confirm a long-period exoplanet from a mono-transit, one must first determine the orbital period. This can be achieved through additional photometric monitoring, spectroscopic follow-up over a complete period, or some combination of the two \citep[e.g.][]{Lendl2020TOI-222:NGTS}. If a second transit is observed some time later around the star with the same depth and shape (but with a time gap), then it becomes a `duo-transit', and the true period is restricted to a set of period aliases that can be efficiently checked through follow-up photometry \citep[e.g.][]{Osborn2022UncoveringTOI-2076}. These duo-transits became increasingly common after the completion of the TESS primary mission in July 2020, as the TESS mission was extended to approximately switch between the northern and southern hemispheres year by year.\footnote{See https://tess.mit.edu/observations/ for full details of the pointing by sector.} However, because typically only one month of data are collected for each region of the sky by TESS every two years, this still misses the majority of long-period transiting exoplanets. Confirming these systems in a more efficient manner necessitates more targeted follow-up with both photometry and spectroscopy \citep{Cooke2020ResolvingMission}.
%A full description of all duotransits found by the NGTS monotransit team in years 1 and 3 of TESS's observations can be found in Hawthorn et al. (submitted)

% Examples
The first transiting planet to be discovered from a mono-transit in TESS data was NGTS-11 b (TOI-1847 b), a warm Saturn in a 35.46-day orbit around a K-type star \citep{Gill2020NGTS-11Event}. The first sign of this planet was a single transit in TESS data; a second transit was found from 79 nights of monitoring observations with the Next Generation Transit Survey \citep[NGTS;][]{Wheatley2018TheNGTS}, which ruled out all but 13 remaining possible periods and allowed the final period to be determined through joint fits to the transit and collected RV data. Since this first discovery, long-period exoplanet discoveries based on single transits in TESS data have multiplied, with over a dozen discovered in the last two years alone \citep{Dalba2020TheCameras,Ulmer-Moll2022TwoB,Grieves2022AnTOI-5542,Ulmer-Moll2023TOI-5678b:HARPS,Gupta2023ATOI-4127,Brahm2023ThreeTESS,Mireles2023TOI-4600Dwarf,Mann2023GiantTOI-2010,Osborn2023TwoiCheops/i}. 

% Paper section overview
This paper describes the discovery and characterisation of a new $\sim$1Gyr old transiting warm Jupiter detected from only one full and a two partial transits, supported by high-precision mass measurements made using CORALIE, the Fiber-fed Extended Range Optical Spectrograph (FEROS), the High Accuracy Radial velocity Planet Searcher (HARPS), and the Carnegie Planet Finder Spectrograph (PFS). The observations of this target and its host star are summarised in Section \ref{observations}, followed by a description of the derivation of stellar parameters, initial photometric and spectroscopic analyses, and the joint modelling in Section \ref{methods}. The final results are presented and discussed in Section \ref{results}, before the paper concludes in Section \ref{conclusions}.

%--------------------------------------------------------------------
\section{Observations} \label{observations}

NGTS-30\,b/TOI-4862\,b was first discovered via a single transit in TESS photometry (Section \ref{tess_obs}), before archival photometry from NGTS (see Section \ref{ngts_obs}) and EulerCam (Section \ref{ecam_obs}) were used to refine the planetary period. Ground-based spectroscopic observations were obtained using the high resolution spectrographs CORALIE (Section \ref{cor_obs}), FEROS (Section \ref{feros_obs}), HARPS (Section \ref{harps_obs}), and PFS (Section \ref{pfs_obs}). Speckle imaging observations to check for other nearby sources were taken using the Southern Astrophysical Research (SOAR) telescope (Section \ref{speckle_obs}).

\subsection{TESS photometry} \label{tess_obs}

\begin{figure*}
\centering
\includegraphics[width=\hsize]{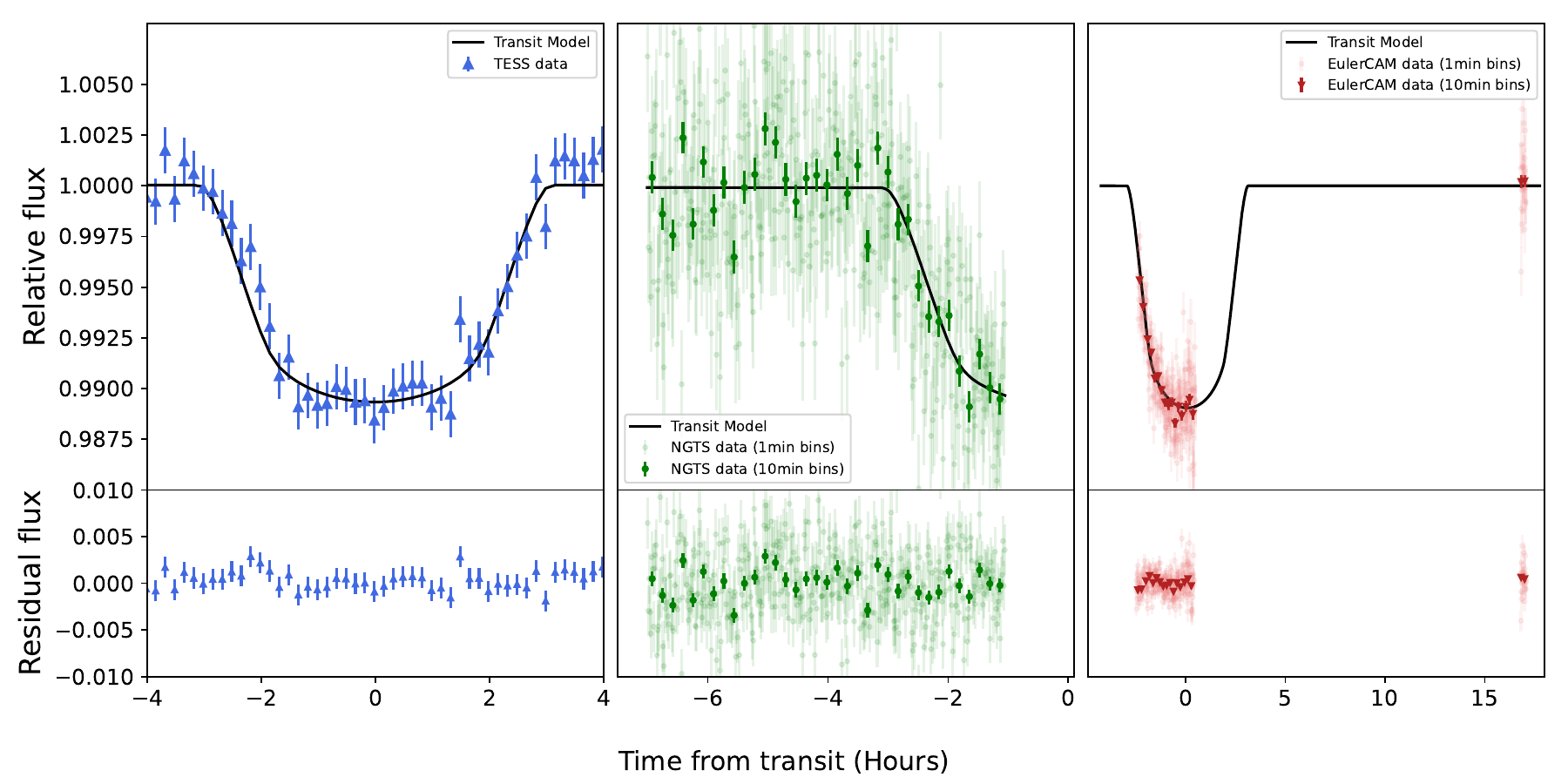}
  \caption{Fitted transits from TESS (left), NGTS (middle), and EulerCam (right) along with their residuals after removing the transit model (lower panels). Because of the high-cadence of the NGTS and EulerCam data, data are shown binned to 1 minute (in lighter colours) and 10 minute bins (darker colours) to align with the TESS data. The joint transit model is over-plotted in black.}
     \label{fig:final_transits}
\end{figure*}

TESS observed TOI-4862 (TIC 322807371) during three different sectors in Years 1, 3, and 5 of its observations: Sector 9 (28 February 2019 to 26 March 2019), Sector 36 (7 March  2021 to 2 April 2021) and Sector 63 (10 March 2023 to 6 April 2023). In Sector 9 it was observed in 30-minute cadence in the TESS Full Frame Images (FFIs), but increased to 10-minute cadence in Sector 36 and 200-second cadence in Sector 63. Because of the long period of this exoplanetary system, only a single transit of TOI-4862\,b was seen in the TESS photometry, occurring at barycentric dynamical time (BJD) 2459287.6 in the Sector 36 observations (n.b. all times are in the barycentric dynamical time frame). This single transit was originally found as part of the Quick-Look Pipeline (QLP) faint star search \citep{Kunimoto2023QLPSearch} resulting in a TOI alert on 21 December 2021. This transit event is shown in Figure \ref{fig:final_transits}.

To rule out possible false positive scenarios for this single transit such as asteroids or background effects, the TESS calibrated Full-Frame-Images were carefully inspected, confirming that the signal came from the star TOI-4862. Furthermore, \textit{Gaia} DR3 results from the surrounding region were reviewed to ensure that the signal could not come from nearby blended stars, planets or eclipsing binaries. No other stars brighter than \textit{Gaia} mag 16.5 were found within the aperture used for TESS photometry. Before being elevated to a TOI by the TESS team, the candidate signal was carefully vetted to rule out clear systematic effects, binaries, astrophysical variability, non-transit shaped signals or signals with clear differences between even and odd transits, as outlined by \citet{Guerrero2021TheMission}. Following elevation, the candidate signal was given the name TOI-4862.01.

TESS light curves of TOI-4862 were obtained from the Mikulski Archive for Space Telescopes,\footnote{https://mast.stsci.edu/portal/Mashup/Clients/Mast/Portal.html} accessing the light curves analysed by the TESS Science Processing Operations Centre (SPOC; \citealt{Jenkins2016TheCenter}, located at NASA Ames Research Center). The TESS-SPOC high level science product FFI light curves were used \citep{Caldwell2020TESSProducts}. The Presearch Data Conditioning Simple Aperture Photometry (PDC\_SAP) fluxes and errors were used for this analysis \citep{Stumpe2012KeplerCurves,Stumpe2014MultiscaleData,Smith2012iKepler/iCorrection}. PDC\_SAP light curves are extracted via aperture photometry from calibrated target pixel files and correct for dilution and systematic trends. The SPOC conducted a transit search in the TESS-SPOC light curve for Sector 36 with a noise-compensating matched filter \citep{Jenkins2002ThePlanets,Jenkins2020KeplerSearch}, which yielded the detection of the single transit of TOI-4862. A limb-darkened model was fitted to the light curve \citep{Li2019iKepler/iSearch} and a suite of diagnostic tests were conducted to make or break confidence in the transiting planet hypothesis \citep{Twicken2018iKepler/iCandidates}, including the difference image centroiding test, which located the host star within 0.9 +/- 2.5 arcsec of the transit source location. We note, however, that only data from Sector 36 data were used in the joint fit as it was the only sector containing a transit. No significant periodic stellar activity was evident in the TESS photometry for this target.

\subsection{NGTS photometry} \label{ngts_obs}

\begin{figure*}
\centering
\includegraphics[width=\hsize,trim=4 4 4 4,clip]{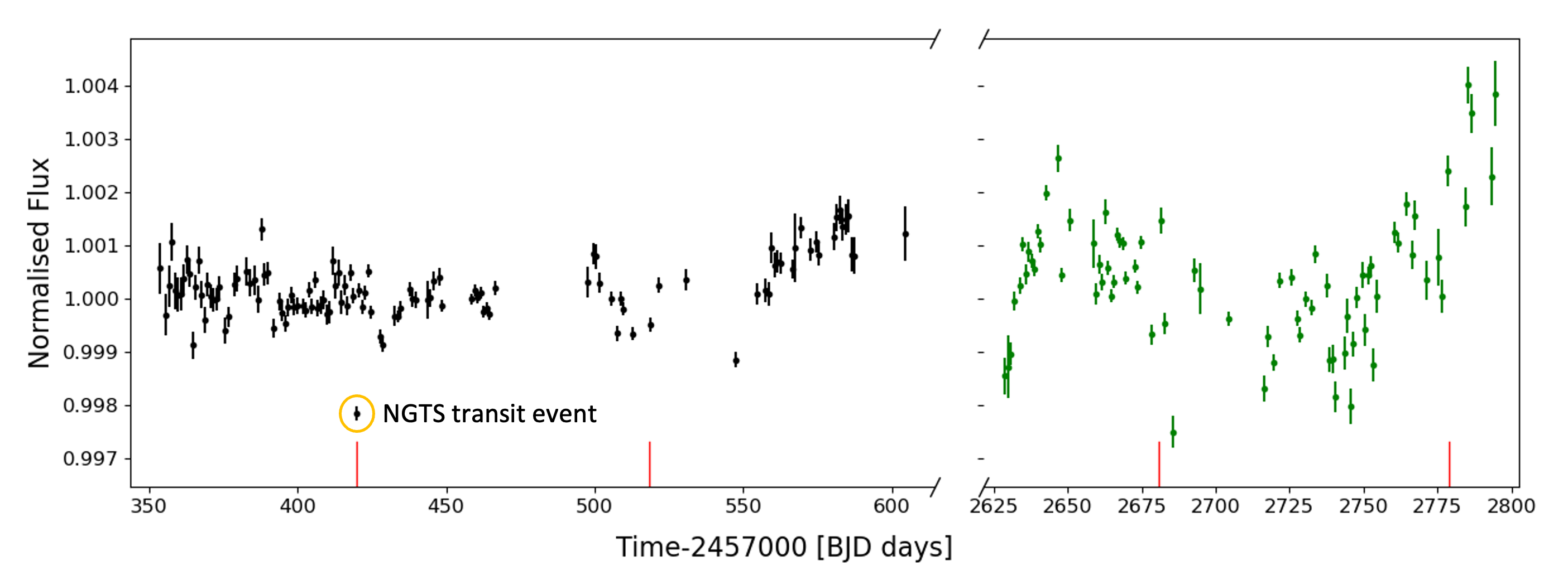}
  \caption{NGTS monitoring plot showing average nightly fluxes for TOI-4862. Archival NGTS data (from before the TESS transit) are plotted in black, and follow-up monitoring is plotted in green with 1-sigma flux errors. Measurements obtained on nights overly affected by noise or weather as well as four monitoring points from different NGTS telescopes with different offsets are not shown. Red lines denote the expected times of transit for NGTS-30\,b/TOI-4862-b, with the first one aligning with the captured NGTS transit (circled). Nightly errors are generally higher in the newer datasets because the typical observation lengths were shorter in this monitoring stage.}
     \label{fig:monitoring}
\end{figure*}

Following the detection of the single transit of TOI-4862\,b by TESS, archival data from the NGTS \citep[][]{Wheatley2018TheNGTS} was searched, in case additional transits of this long-period planet had been independently observed. This process revealed that NGTS had 135 nights of existing archival photometry for TOI-4862\,b (see Figure \ref{fig:monitoring}), including an ingress of the same depth and shape as the single transit in TESS from 2016 (BJD 2457419.9, see Figure \ref{fig:final_transits}). NGTS comprises 12 automated 20 cm telescopes located at the Paranal site of the European Southern Observatory (ESO). Data are typically taken with one telescope at a time in blind-survey mode, but this can be increased to several telescopes for important transits \citep[e.g. see][]{Bryant2020SimultaneousWASP-166b,Smith2020ShallowTelescopes}. Specially designed to search for transiting exoplanetary signals from the ground, NGTS is capable of high-precision photometry that matches TESS (in regard to precision) for Tmag > 12 for a single telescope and Tmag > 9 for multiple telescopes observing concurrently \citep{Bryant2020SimultaneousWASP-166b}. All NGTS data are acquired with exposure times of 10 seconds and a cadence of 13 seconds. Magnitude data were extracted from the telescope using the standard NGTS aperture photometry extraction pipeline and systematics removal detailed in \citet{Wheatley2018TheNGTS}. Following extraction of the light curves, transits were searched for using the template-matching technique described in \citet{Gill2020NGTS-11Event}.

As the archival NGTS data included only a single ingress for this target, additional observations were performed between February 2022 and July 2023 in order to search for additional transits and constrain the period further. This resulted in another 125 nights of data, for a total of 260 nights of data between November 2015 and July 2023 for TOI-4862. Although no additional transits were found in this search, the resulting extensive dataset allowed the majority of the possible period aliases for the planet candidate to be ruled out, as discussed in section \ref{method_phot}. All NGTS monitoring data are plotted in Figure \ref{fig:monitoring}, excluding those nights that were hampered by poor observing conditions and a further four nights that were taken on different NGTS telescopes.

\subsection{EulerCam photometry} \label{ecam_obs}

In the final stages of preparing this paper, the opportunity arose to observe an additional transit of TOI-4862 b with EulerCam on the night of 27 February 2024, aligning with the most likely period alias of 98.3 days. As all points from this observation where during the transit, an additional follow-up observation was also carried out the following night (28 February 2024) to confirm the out of transit flux baseline. %

EulerCam is a 4k $\times$ 4k CCD detector installed in the 1.2 m \textit{Leonhard Euler }Telescope at ESO's La Silla Observatory \citep{Lendl2012}.  The observations were acquired in the NGTS filter with an exposure time of 30 seconds and 0.1 mm de-focus. The raw full-frame images are corrected for bias, over-scan and flat field using the standard EulerCam reduction pipeline. The aperture photometry is performed by placing circular apertures of varying radii on the target and a few reference stars.  The exact placement of these apertures is performed using the astrometric solution of stars in the field-of-view \citep{astrometry_net}.  The optimal aperture size and reference stars are selected by minimising the root mean square scatter of the light curve, which in this case was 40 pixels.

This new observation showed a partial transit with the same depth and shape as the TESS and NGTS transits (as shown in Figure \ref{fig:final_transits}), confirming that the 98.3 day period alias was the true period.

\begin{table}
\caption{RV data for TOI-4862.}             % title of Table
\label{rv_table}      % is used to refer this table in the text
\centering                          % used for centering table
\begin{tabular}{c c c c}        % centered columns (4 columns)
\hline\hline                 % inserts double horizontal lines
Time & RV        & RV Error  & Instrument \\
BJD  & [ms$^{-1}$] & [ms$^{-1}$] &  - \\    % table heading 
\hline                        % inserts single horizontal line
2459598.79120 & -2.23 & 28.607552 & CORALIE \\
2459605.74120 & -15.54 & 31.680687 & CORALIE \\
2459611.69435 & -75.24 & 38.410280 & CORALIE \\
   ...        &         ...  & ...       & ... \\
2460074.59522 &      0.00    &  2.04     &      PFS \\
2460122.49911 &         -37.28   &      2.05     &      PFS \\ 
2460122.51344 &         -40.19   &      2.08     &      PFS \\
\hline                                   %inserts single line
\\
\multicolumn{4}{l}{Note: The full table of RV data is available at the CDS.}\\
\end{tabular}
\end{table}

\subsection{TRES reconnaissance spectroscopy} \label{tres}

Two reconnaissance spectra were obtained with the Tillinghast Reflector Echelle Spectrograph \citep[TRES;][]{Furesz2008DesignClusters} located at the \textit{Fred Lawrence Whipple }Observatory (FLWO) on 16 February 2022 and 15 March 2022. TRES is a fibre-fed echelle spectrograph operating in the 3900-9100 Angstrom wavelength range with a spectral resolution of R=44,000. The spectra were extracted and reduced following procedures outlined in \citet{Buchhave2010HAT-P-16b:ORBIT} and used to derive stellar parameters using the Stellar Parameter Classification (SPC) tool \citep[][]{Buchhave2014ThreeMetallicities}. SPC derives an effective temperature, surface gravity, metallicity, and rotational velocity by cross-correlating the observed spectra against Kurucz atmospheric models \citep{Kurucz1992ModelSynthesis}. The derived stellar parameters were: Teff = 5479 $\pm$ 50\,K, logg = 4.60 $\pm$ 0.10, [m/H] = 0.03 $\pm$ 0.08, vsini = 3.3 $\pm$ 0.5 km/s and a signal to noise per resolution element of SNRe = 28.1. We note, however, that because only two spectra were obtained by TRES, these data are not included in the full RV analysis.

\subsection{CORALIE spectroscopy} \label{cor_obs}

\begin{figure}
\centering
\includegraphics[width=\hsize]{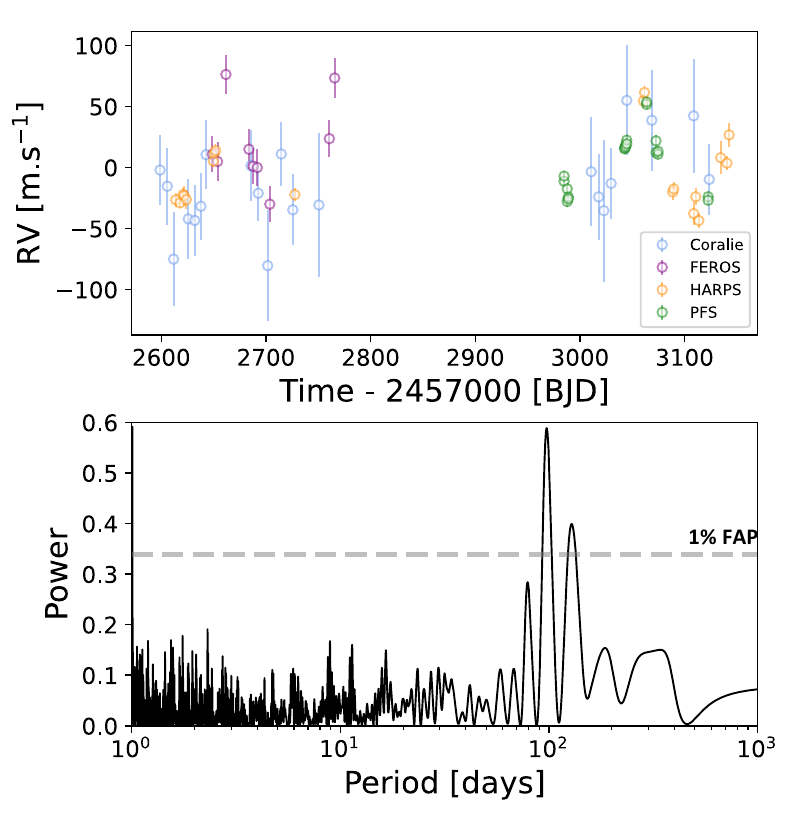}
  \caption{Overview of collected RV data \textit{Top:} All collected RV observations for TOI-4862. CORALIE points are shown in blue, FEROS in purple, HARPS in orange, and PFS in green. All points are shown as empty circles, with their 1$\sigma$ errors shown as straight lines. Inter-instrument offsets have been removed. \textit{Bottom:} Lomb-Scargle periodogram for the collected RV observations. The 1\% false alarm probability (FAP) is shown by the dotted grey line. The highest peak corresponds to a period of about 97.4 days.}
     \label{fig:rv_obs}
\end{figure}

TOI-4862 was added to an existing programme on the CORALIE spectrograph \citep{Queloz2000The86} to follow up on long-period transiting exoplanet candidates from TESS. CORALIE is a fibre-fed echelle spectrograph installed at the Swiss 1.2m \textit{Leonhard Euler }Telescope at the La Silla site of the ESO in Chile. The CORALIE spectrograph uses two fibres for fibre injection: the first is dedicated to observing the target, and the second can focus on either light from a Fabry-Pérot etalon to perform simultaneous drift calibration or the sky to allow for background subtraction. In this case, the Fabry-Pérot etalon was used. Spectra were extracted from the detector using the standard CORALIE calibration reduction pipeline, before RV measurements were determined through cross-correlation with a binary G2 mask \citep{Pepe2002TheVII}. In total 20 CORALIE spectra of TOI-4862 were taken between 19 January 2022 and 28 June 2023, with exposure times of 30-45\,min depending on observing conditions. The extracted RV measurements had an overall dispersion of <70 m.s$^{-1}$, providing good evidence that the observed transit signal came from a planetary-mass object. All CORALIE RV measurements are plotted in Figure \ref{fig:rv_obs} and can be found in Table \ref{rv_table}.

\subsection{FEROS spectroscopy} \label{feros_obs}

Following the beginning of the CORALIE observations, nine additional spectra were obtained using the FEROS spectrograph. FEROS is a high resolution spectrograph installed on the 2.2m telescope at the La Silla site of the ESO \citep{Kaufer1999CommissioningLa-Silla}. FEROS has a spectral resolution of 48,000, corresponding to 2.2 pixels of 15$\mu$m, and covers the wavelength range of 3600-9200\AA. The eight FEROS spectra were collected between 10 March 2022 and 5 July 2022 as part of the ongoing programme to follow up long-period planets (PI: Hobson; ESO Program ID 0108.A-9003(A) and 0109.A-9003(A)). All data were taken using an exposure time of 1500s. Radial velocity measurements were then extracted from each spectrum using the cross-correlation technique, according to the CERES pipeline outlined in \citet{Brahm2017CERES:Spectra}. These RV measurements are summarised in Table \ref{rv_table} and plotted in Figure \ref{fig:rv_obs}.

\subsection{HARPS spectroscopy} \label{harps_obs}

In order to more precisely constrain the mass of NGTS-30\,b/TOI-4862\,b, 19 spectra were obtained for TOI-4862 on the HARPS spectrograph \citep{Pepe2002TheVII}. HARPS is a fibre-fed spectrograph installed on the 3.6m telescope at La Silla. It has a spectral resolution of $\sim$115,000 and covers the wavelengths of 3800-6900\AA. The observations of TOI-4862 were taken as part of ongoing long-period transiting giants programmes (108.22L8, 109.23J8 \& 111.250B, PI: Ulmer-Moll). TOI-4862 was observed in the high-accuracy mode between 3 February 2022 and 17 July 2023, with exposure times of 1800s. All HARPS data were reduced using version 3.0 of the standard HARPS reduction pipeline \citep{Lovis2007AVisible}. Radial velocities were then extracted from these spectra through cross-correlation with a G2 mask. These RV measurements are summarised in Table \ref{rv_table} and plotted in Figure \ref{fig:rv_obs}.

\subsection{PFS spectroscopy} \label{pfs_obs}

Twenty additional RV measurements of TOI-4862 were taken concurrently by the Carnegie PFS (\citealt{Crane2006TheSpectrograph,Crane2008TheReport,Crane2010TheCommissioning}). PFS is a high-precision echelle spectrograph attached to the 6.5 metre Magellan Clay telescope at Las Campanas Observatory in Chile. It has a spectral resolution of $\sim$130,000 and covers the 3900-7340\AA\,spectral window. Wavelength calibration is carried out using an iodine absorption cell, which also allows for the characterisation of the instrumental profile. Spectra were reduced using the standard PFS reduction pipeline \citep{Butler1996AttainingS-1,Crane2006TheSpectrograph} and RV measurements were extracted using a custom IDL pipeline. TOI-4862 was observed by PFS between 9 February 2023 and 27 June 2023, for a total of 20 individual measurements. All PFS RV measurements are summarised in Table \ref{rv_table} and plotted in Figure \ref{fig:rv_obs}.

\subsection{SOAR speckle imaging} \label{speckle_obs}

\begin{figure}
\centering
\includegraphics[width=\hsize]{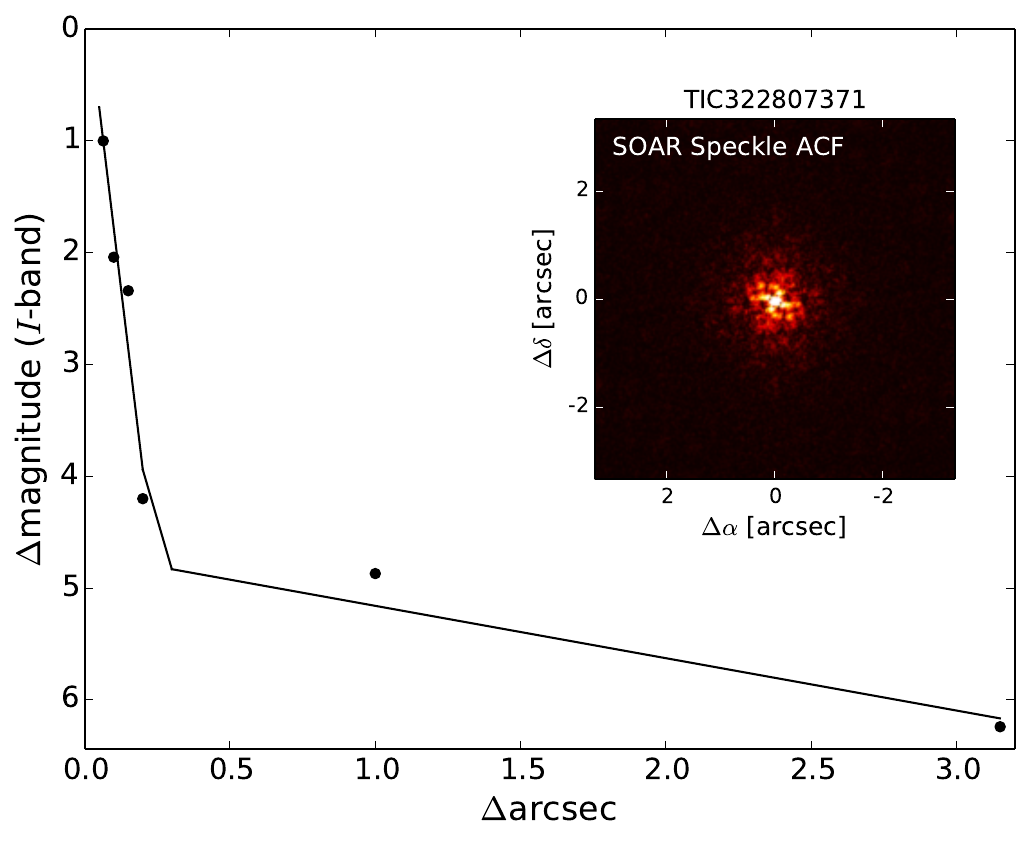}
  \caption{SOAR speckle-imaging observations of TOI-4862. No additional sources were observed within 3$\arcsec$ of the target.}
     \label{fig:SOAR_obs}
\end{figure}

High-angular resolution imaging is needed to search for nearby sources that can contaminate the TESS photometry (resulting in an underestimated planetary radius) or be the source of astrophysical false positives such as background eclipsing binaries. A search for stellar companions to TOI-4862 was done with speckle imaging on the 4.1 m SOAR telescope \citep{Tokovinin2018TenSOAR} on 20 March 2022 UT in the Cousins I-band, a band-pass similar to that of a TESS camera. This observation was sensitive to a 5.0-magnitude fainter star at an angular distance of 1 arcsec from the target, but no nearby stars were detected within 3$\arcsec$ of TOI-4862. More details of the observations within the SOAR TESS survey are available in \citet{Ziegler2019SOARCompanions}. The 5$\sigma$ detection sensitivity and speckle auto-correlation functions from the observations are shown in Figure \ref{fig:SOAR_obs}. 

%-----------------------------------------------------------------
\section{Data analysis} \label{methods}

\subsection{Stellar parameter determination} \label{method_star}

The collected HARPS spectra were used to determine the properties of the host star TOI-4862, co-adding the 15 spectra by correcting them to a common wavelength scale and combining them to a single spectrum with signal-to-noise of $\sim$90. Following \citet{Gill2020NGTS-11Event}, the \texttt{ISPEC} \citep{Blanco-Cuaresma2014DeterminingISpec} and \texttt{SPECTRUM} \citep{Gray1999SPECTRUM:Program} codes were used to model a grid of synthetic spectra based on MARCS model atmospheres \citep{Gustafsson2008AStars}, version 5 of the \textit{Gaia}-ESO Survey atomic line lists and solar abundances from \citet{Asplund2009TheSun}. Based on these models, the stellar effective temperature, $T_{\mathrm{eff}}$, and stellar surface gravity, log $g$, were calculated from the H$\alpha$, NaI D, and MgI b lines, while the metallicity, Fe/H, and rotational broadening, $v\sin i$, were determined from the individual FeI and FeII lines. 

\begin{figure}
\centering
\includegraphics[width=\linewidth]{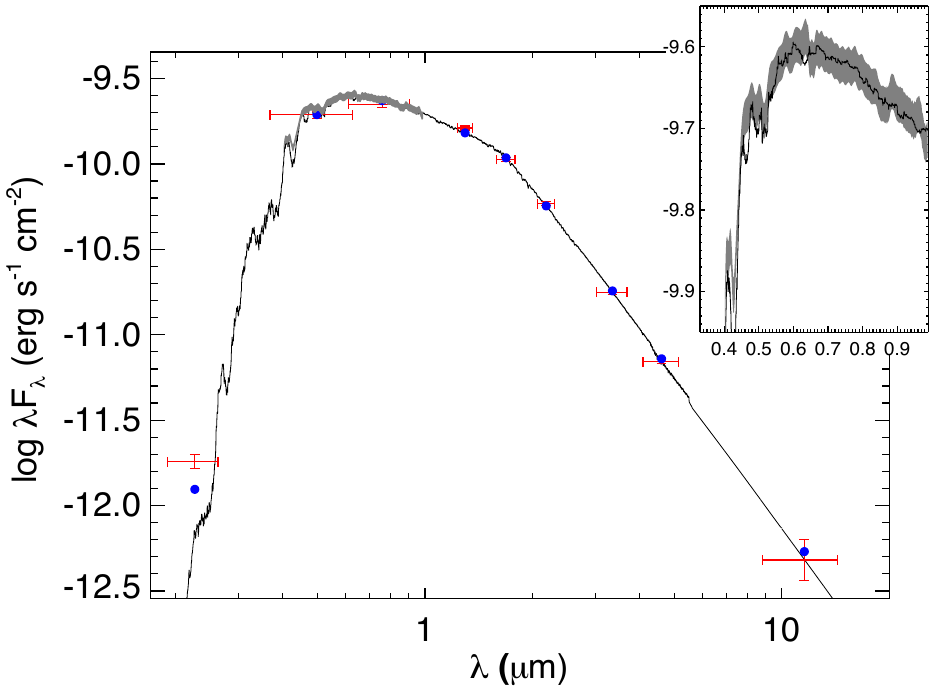}
\caption{Spectral energy distribution of TOI-4862. Red symbols represent the observed photometric measurements, and the horizontal bars represent the effective width of the passband. Blue symbols are the model fluxes from the best-fit PHOENIX atmosphere model (black). The absolute flux-calibrated {\it Gaia\/} spectrum is shown as a grey swathe in the inset figure. We note that the discrepancy between the blue points and PHOENIX atmosphere model at the shortest wavelengths is due to the SED being so steep in the UV that the nominal band-pass effective wavelength ends up being different from the flux-weighted effective wavelength \label{fig:sed}}
\end{figure}

An analysis of the broadband spectral energy distribution (SED) of the star was performed together with the {\it Gaia\/} DR3 parallax, in order to determine an empirical measurement of the stellar radius \citep{Stassun:2016,Stassun:2017,Stassun:2018}. The $JHK_S$ magnitudes were sourced from 2MASS, the W1--W3 magnitudes from WISE, the $G_{\rm BP}$ and $G_{\rm RP}$ magnitudes from {\it Gaia}, and the near-UV magnitude from GALEX. The absolute flux-calibrated {\it Gaia\/} spectrum was also utilised. Together, the available photometry spans the full stellar SED over the wavelength range 0.2--10~$\mu$m (see Figure~\ref{fig:sed}). 

A fit using PHOENIX stellar atmosphere models \citep{Husser:2013} was performed, adopting from the spectroscopic analysis the effective temperature ($T_{\rm eff}$), metallicity ([Fe/H]), and surface gravity ($\log g$). The extinction $A_V$ was fitted for, limited to the maximum line-of-sight value from the Galactic dust maps of \citet{Schlegel:1998}. The resulting fit (Figure~\ref{fig:sed}) has $A_V = 0.11 \pm 0.06$. Integrating the (un-reddened) model SED gives the bolometric flux at Earth, $F_{\rm bol} = 3.628 \pm 0.085 \times 10^{-10}$ erg~s$^{-1}$~cm$^{-2}$. Taking the $F_{\rm bol}$ together with the {\it Gaia\/} parallax directly gives the bolometric luminosity, $L_{\rm bol} = 0.663 \pm 0.016$~L$_\odot$. The Stefan-Boltzmann relation then gives the stellar radius, $R_\star = 0.913 \pm 0.029$~R$_\odot$. In addition, the stellar mass was estimated using the empirical relations of \citet{Torres:2010}, giving $M_\star = 0.94 \pm 0.06$~M$_\odot$. 

\begin{figure}
\centering
\includegraphics[width=\linewidth]{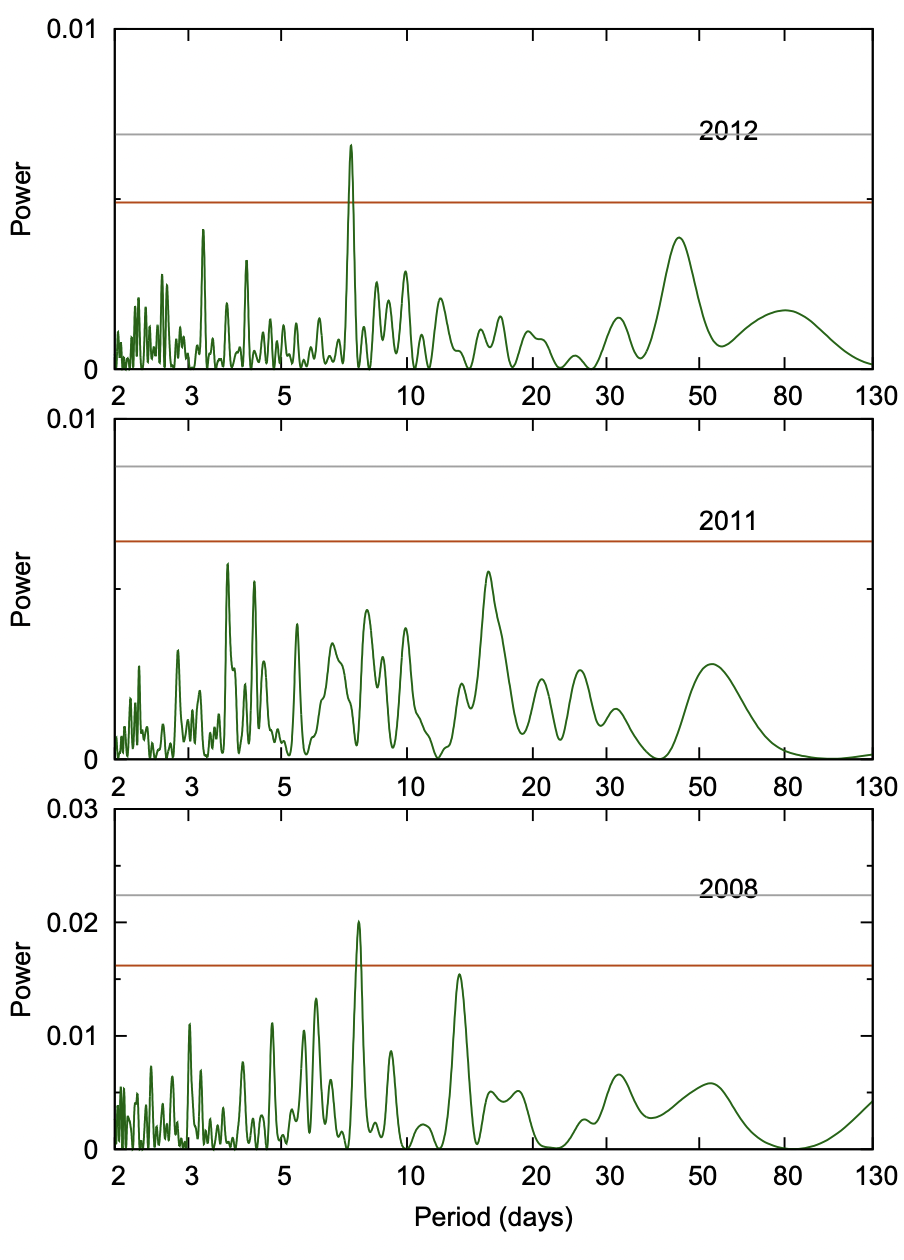}
\caption{WASP rotation analysis: Lomb-Scargle periodograms were run for three separate years of WASP photometry, in 2012, 2011, and 2008 (from top to bottom). 1\% and 10\% false alarm probability levels (estimated with methods from \citealt{Maxted2011WASP-41b:Star}) are shown as grey and red lines, respectively. The same $\sim$ 7.5 day rotation signal is marginally detected in 2012 and 2008, while the highest peak in the 2011 data is at 3.73 days, which could be a harmonic of the 7.5 day signal.}
\label{fig:wasp}
\end{figure}

Finally, the projected stellar rotation period, $P_{\rm rot} / \sin i$, was estimated from the spectroscopic $v\sin i$ together with $R_\star$, giving $13.2 \pm 3.0$~d. From the empirical gyrochronology relations of \citet{Mamajek:2008}, the system age can then be estimated to be $1.1 \pm 0.4$~Gyr. Curiously, although no reliable rotation period was retrieved from the NGTS monitoring data (nor in the TESS sectors), archival longer-term photometric monitoring by the Wide Angle Search for Planets (WASP) survey \citep{Pollacco2006TheCameras} marginally detected a shorter period of 7.5 $\pm$ 0.3 days (see Figure \ref{fig:wasp}), which matches the value from $v\sin i$ if $i \sim 35^{\circ}$. This would suggest that the planet's orbit is misaligned with the stellar spin axis, and would also lead to a gyrochronological age even younger than 1.1 Gyr. Given the young age of this target, further observations to more reliably determine the rotation period and (mis)alignment of the star would be highly valuable.

All stellar properties extracted from this analysis can be found in Table \ref{stellar_table} and are in good agreement with both those in version 8 of the TESS Input Catalogue \citep{Stassun2019TheList} and those extracted from the TRES spectra on ExoFOP.\footnote{https://exofop.ipac.caltech.edu/tess/target.php?id=322807371}

\begin{table}
\caption{Stellar names, properties, and photometric magnitudes.}             % title of Table
\label{stellar_table}      % is used to refer this table in the text
\centering                          % used for centering table
\begin{tabular}{l c c}        % centered columns (3 columns)
\hline\hline                 % inserts double horizontal lines
Parameter  & Value & Source \\    % table heading 
\hline                        % inserts single horizontal line
Names\\
TOI & TOI-4862 & TESS \\
TIC ID & 322807371 & TESS \\
2MASS ID & J11345152-2436189 & 2MASS \\
\textit{Gaia} DR3 ID & 3533836281547278848 & \textit{Gaia} \\ \\

Astrometric properties \\
RA  & 11:34:51.57   &   \textit{Gaia}  \\
Dec & -24:36:19.75  &   \textit{Gaia} \\ 
pmRA [mas/yr] & 23.7449 ± 0.0119 &     \textit{Gaia}\\
pmDec [mas/yr] & -49.3001 ± 0.0109 & \textit{Gaia}\\
Parallax [mas] & 4.1305 ± 0.0157 & \textit{Gaia}\\
Distance [pc] & 233.9 ± 1.2 & \textit{Gaia}\\ \\

Magnitudes\\
TESS [mag] & 11.6707 ± 0.006 & TESS\\
V [mag] & 12.537 ± 0.214 & Tycho\\
B [mag] & 12.221 ± 0.315 & Tycho\\
G [mag] & 12.1646 ± 0.0001 & \textit{Gaia}\\
J [mag] & 10.943 ± 0.023 & 2MASS\\
H [mag] & 10.601 ± 0.027 & 2MASS\\
Ks [mag] & 10.501 ± 0.022& 2MASS\\
WISE1 [mag] & 10.475 ± 0.023 & WISE\\
WISE2 [mag] & 10.512 ± 0.02 & WISE\\
WISE3 [mag] & 10.491 ± 0.085 & WISE\\
WISE4 [mag] & 8.297 & WISE\\ \\

Bulk properties\\
$T_{\mathrm{eff}}$ [K]  & 5455 $\pm$ 80       & Sect. \ref{method_star}\\
Fe/H [dex]            & -0.03 + 0.08          & Sect. \ref{method_star}\\
log g [cm.s$^{-2}$]       & 4.3 $\pm$ 0.5     & Sect. \ref{method_star}\\
$v \sin i$ [km.s$^{-1}$]  & 3.5 $\pm$ 0.8     & Sect. \ref{method_star}\\
M$_\star$ [M$_\odot$]     & 0.94 $\pm$ 0.06   & Sect. \ref{method_star}\\
R$_\star$ [R$_\odot$]     & 0.913 $\pm$ 0.029 & Sect. \ref{method_star}\\
Age [Gyr]                 & 1.1 $\pm$ 0.4     & Sect. \ref{method_star}\\
\hline                                   %inserts single line
\\
\end{tabular}
\textbf{References:} TESS \citep{Stassun2019TheList}; 2MASS \citep{Skrutskie2006The2MASS}; \textit{Gaia} DR3 \citep{GaiaCollaboration2016TheMission}; Tycho \citep{Hog2000TheStars}; WISE \citep{Wright2010THEPERFORMANCE}.
\end{table}

\subsection{Photometric analysis} \label{method_phot}

\begin{figure}
\centering
\includegraphics[width=\hsize,trim={0.1 0 0 1cm},clip]{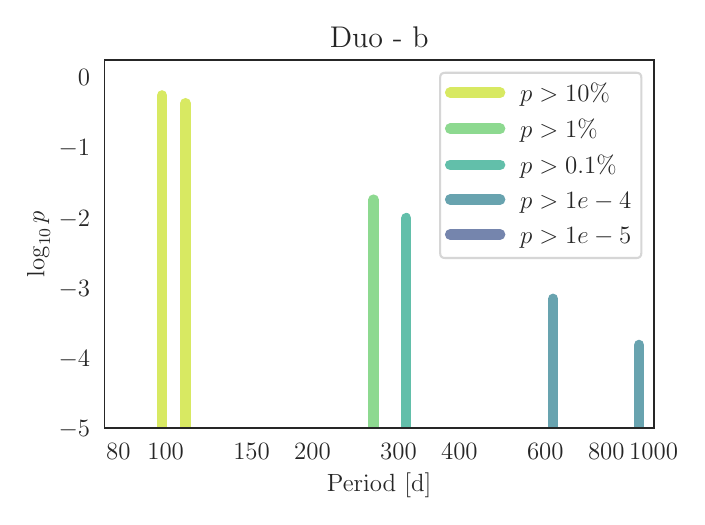}
  \caption{Most likely period aliases based on the TESS and NGTS photometric fit from the \texttt{MonoTools} package \citep{Osborn2022UncoveringTOI-2076}. The most favoured remaining period is 98.3 days, aligning with the favoured period from the RV-only period search (section \ref{method_rv}). The y-axis shows the log likelihood.}
     \label{fig:monotools}
\end{figure}

As only one transit of NGTS-30\,b/TOI-4862\,b was found in the TESS data and only an ingress was recovered in the NGTS data, the true period of the planet was initially unknown. In order to determine the separation of the TESS and NGTS transits, the two transits were fit simultaneously with a basic \texttt{batman} transit model \citep{Kreidberg2015Batman:Python}, utilising the transit depth from the TESS transit and wide uniform priors on the period. This resulted in a maximum period of 1867.68$^{+0.19}_{-0.16}$ days. However, because only two transits existed for this target, many possible period aliases remained, determined from the equation $ P_{\mathrm{n}} = 1867.7/n$, where $n$ is an integer. To test each of these aliases, the archival and new photometric data were folded by each alias down to n = 52 (P= 35.9 days), and those aliases with clearly flat NGTS data were discarded. Given the length of the transit in the TESS data and extensive monitoring by NGTS, any periods shorter than 36 days were already implausible. As aliases were ruled out, the NGTS follow-up was directed to focus solely on epochs of aliases that remained. Following this photometric follow-up, only 6 aliases remained plausible, excluding the original 1867-day period. The remaining period aliases were analysed using the \texttt{MonoTools} package \citep{Osborn2022UncoveringTOI-2076}, which performs a Bayesian comparison of the most likely period aliases based on the shape of the transits. This analysis determined that the two shortest remaining periods, 98.3 and 109.9 days, were most favoured as shown in Figure \ref{fig:monotools}. As this manuscript was in the final stages of completion, an additional opportunity arose to observe the most likely alias of 98.3 days with EulerCam on the  Swiss 1.2m \textit{Leonhard Euler} Telescope.  This additional observation on 27 February 2024 revealed a transit signal of the same depth as that in TESS and NGTS, confirming that the 98.3 day period was correct through photometric data alone.

\subsection{Radial velocity analysis} \label{method_rv}

\begin{figure}
\centering
\includegraphics[width=\hsize]{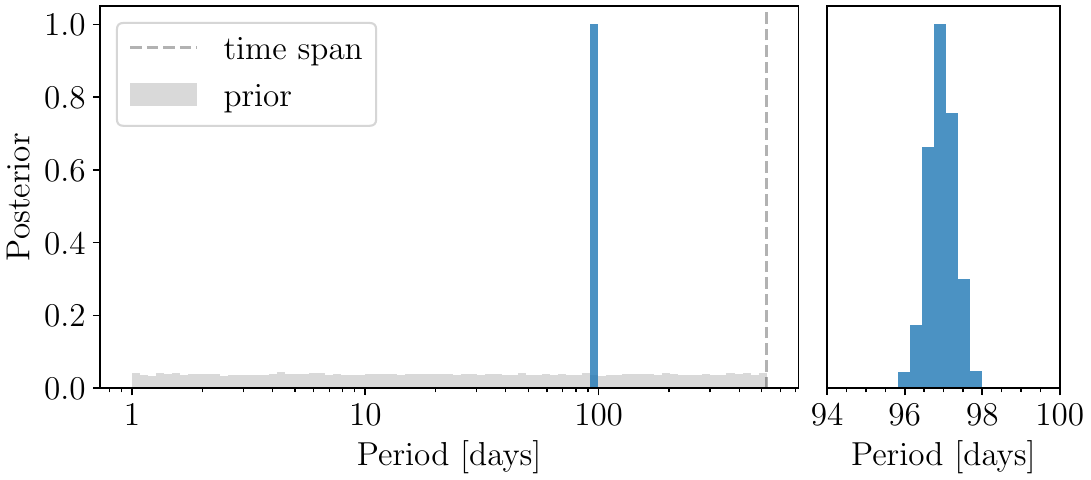}
  \caption{Posterior distribution (blue) for the orbital period from the \texttt{kima} fit of only RV data. The posterior distribution's highest peak corresponds to a period of 97 days. The prior distribution is shown as an extended grey box near the bottom of the plot and is log-uniform from 1 day to the time span of the data at $\sim$500 days, as shown by the dotted grey line.}
     \label{fig:kima2}
\end{figure}

An independent fit of only the RV data was also completed ahead of the joint analysis, using the \texttt{kima} package (Faria et al. 2018) to search for periodic signals in the combined data from all four spectroscopic instruments. The \texttt{kima} package uses Bayesian inference to model a sum of Keplerian signals, and also allows the evidence for different numbers of planets to be easily compared. As an initial test, the number of planets was allowed to vary as a free parameter between zero to two. The \texttt{kima} model was run for 100,000 steps with the new level interval set to 3000. This analysis revealed that the one-planet model was the most likely, with the one-planet model (ln(Z) = -290.5, where Z denotes the likelihood) several times more likely than the zero-planet model (ln(Z) = -338.6) and about eight times more likely than the two-planet model (ln(Z) = -292.4). In order to determine the most likely period from the RV data alone, a second \texttt{kima} run was completed with a set one-planet model, which suggested that the most likely period was 97 days (as shown in Figure \ref{fig:kima2}), in good agreement with the most likely period from the photometric-only fit completed in section \ref{method_phot}. Notably, there was no evidence for the second most likely period (109.9 days) from the spectroscopic fit, with the long baseline of the HARPS data effectively ruling out this period alias. 

As an independent check, a Lomb-Scargle period search \citep{Lomb1976Least-squaresData,Scargle1982StudiesData} was also carried out, giving a wide peak of maximum likelihood around 97.4 days (see Figure \ref{fig:rv_obs}). This gave good confidence that the 98.3 day period was the correct period for NGTS-30\,b/TOI-4862\,b, allowing this to be used as a prior in the joint fit below. The RV time series for these data is shown in Figure \ref{fig:rv_obs}, alongside the results from the generalised Lomb-Scargle periodogram run on the RV data alone.

% \begin{figure}
% \centering
% \includegraphics[width=\hsize]{Figures/0-2_planets_plot1.png}
%   \caption{Comparative evidence for 0-2 planets from the radial-velocity only fit.}
%      \label{fig:kima1}
% \end{figure}

\subsection{Global modelling}

Global modelling for NGTS-30\,b/TOI-4862\,b was completed using the \texttt{juliet} package \citep{Espinoza2019Juliet:Systems}. \texttt{juliet} combines transit fitting using \texttt{batman} \citep{Kreidberg2015Batman:Python}, RV fitting using \texttt{radvel} \citep{Fulton2018RadVel:Toolkit}, Gaussian process modelling with \texttt{george} \citep{celerite2:foremanmackey17} and \texttt{celerite} \citep{celerite2:foremanmackey17}, and nested sampling in order to allow it to efficiently and simultaneously fit data from multiple photometric and spectroscopic datasets. Several different nested sampling algorithms are included in the \texttt{juliet} package, but in this case \texttt{dynesty} \citep{Speagle2020Dynesty:Evidences} was used to conduct the global modelling. As in the \texttt{kima} fit, different RV instruments can be included in the fit by also fitting for the RV offsets between the different datasets.

Alongside the shared parameters such as period, eccentricity, argument of periastron and stellar density, additional terms included in the photometric part of the global model were the impact parameter, radius ratio, mid-transit time ($T_0$), mean flux offsets and limb-darkening coefficients for each instrument (following the \citet{exoplanet:kipping13} parameterisation of q1 and q2), while the spectroscopic part of the model included extra parameters for the RV semi-amplitude and individual spectrograph offsets. Parameters with prior values calculated from the photometric, spectroscopic and/or stellar parameter analysis such as the orbital period, $T_0$, stellar density, and the mean flux values, were fit with normal distributions around the calculated mean and inflated standard deviations (roughly five times the calculated standard deviation), while unknown parameters were typically allowed to vary across wide but realistic parameter spaces in uniform or log-uniform distributions. A reasonably narrow normal distribution ($\sigma$ = 1.5 days) was used as a prior for the 98.3 day period to avoid the other aliased periods ruled out by photometric monitoring but was kept wide enough to account for the slight difference between this photometrically determined period and those from RV measurements alone ($\sim$97 days). Meanwhile because neither the NGTS nor EulerCam data contained a full transit, $T_0$ was taken to be the centre of the TESS single transit. Fitting the stellar density was chosen instead of the scaled semi-major axis (a/R$_*$). The stellar mass and radius were based directly on the conducted stellar analysis in section \ref{method_star}, allowing a precise prior to be placed on this parameter. Because the TESS PDCSAP light curves automatically correct for for dilution as part of the extraction pipeline, the dilution factor for TESS was not included in the fit. Likewise, because of the much smaller pixel sizes of the NGTS and EulerCam detectors, these light curves were not found to be affected by dilution either, so the dilution term was also fixed to 1.0 for these datasets.

Priors on the limb-darkening coefficients for TESS, NGTS, and EulerCam were calculated using the LDCU code\footnote{https://github.com/delinea/LDCU}, a modified version of the python routine implemented by \citet{Espinoza2015LimbParameters} that computes the limb-darkening coefficients and their corresponding uncertainties using a set of stellar intensity profiles accounting for the uncertainties on the stellar parameters. The stellar intensity profiles are generated based on two libraries of synthetic stellar spectra: ATLAS \citep{Kurucz1979ModelStars} and PHOENIX \citep{Husser:2013}.

As none of the photometric data display significant correlated noise in their datasets around the transits, a periodic Gaussian process term was not included in this fit, simply fitting additional `jitter' terms in quadrature to both the photometric and spectroscopic data in case the error bars on any of the instruments had been underestimated. 

The \texttt{juliet} model was run using the \texttt{dynesty} nested sampling method with 1100 live points, continuing until the estimated uncertainty in the log evidence was smaller than 0.1. All values and distributions used for the fitted parameters can be found in Table \ref{joint_priors_table}, while the results from the combined fit are displayed in Table \ref{results_table}.

\begin{figure}
\centering
\includegraphics[width=\hsize]{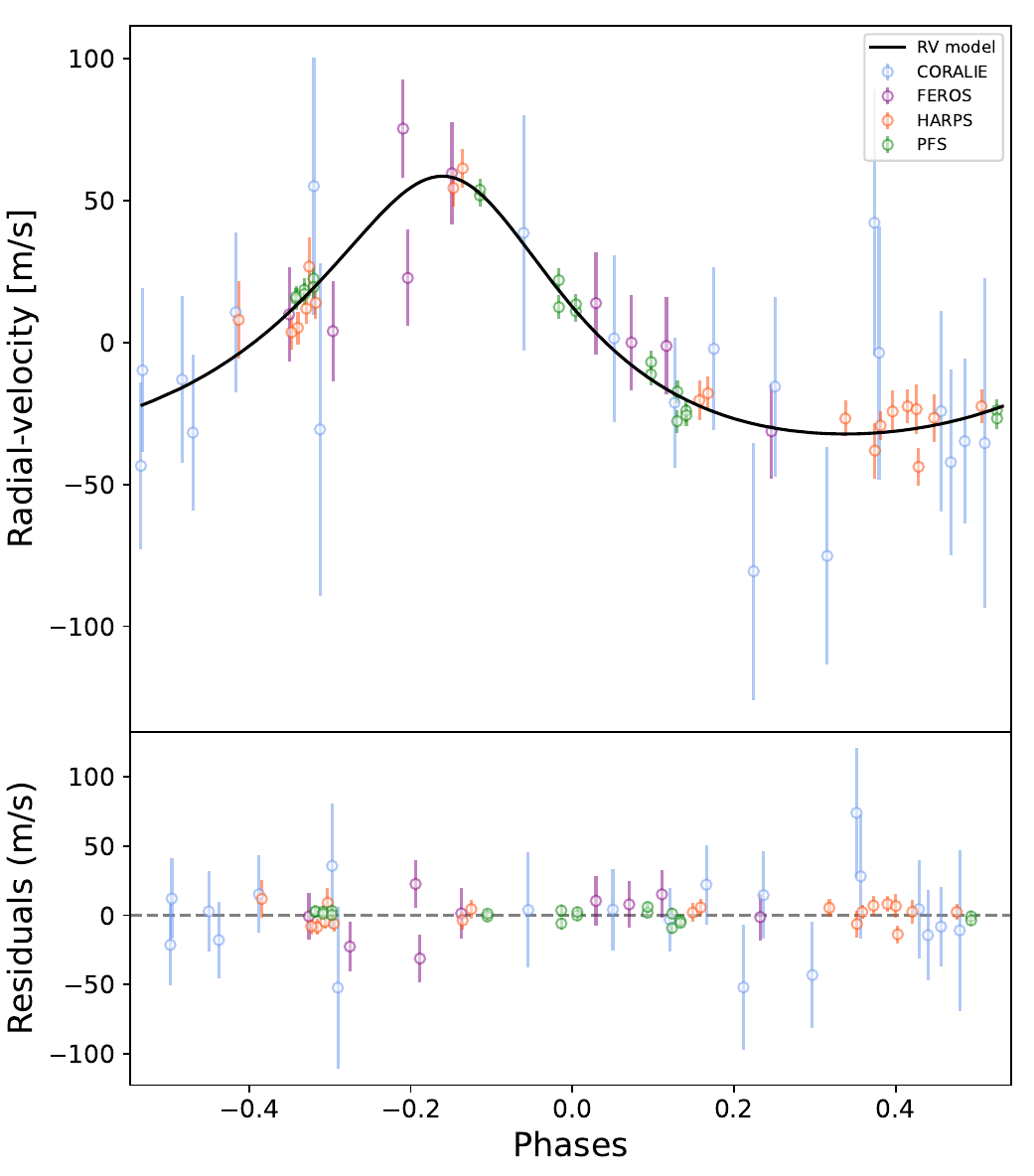}
  \caption{Phase-folded RV data for the  NGTS-30/TOI-4862 system. \textit{Top:} Collected RV data with the joint planet model over-plotted in black. \textit{Bottom}: RV residuals after subtracting the RV model. For clarity, a dashed grey line is plotted at zero on the residuals plot. The eccentricity of NGTS-30\,b/TOI-4862\,b's orbit is clearly shown by the asymmetry of the RV model.}
     \label{fig:rv_fit}
\end{figure}

\begin{table}[htb!]
\centering                          % used for centering table
\caption{Final results for the exoplanet NGTS-30\,b/TOI-4862\,b.}             % title of Table
\label{results_table}      % is used to refer this table in the text
    \begin{tabular}{l c}        % centered columns (3 columns)
    \hline\hline                 % inserts double horizontal lines
    Parameter  & Value \\    % table heading 
    \hline                        % inserts single horizontal line
    Fitted Parameters \\
    Orbital Period [days] & 98.29838$^{+0.00010}_{-0.00009}$  \\
    Time of transit $T_0$ [BJD days] & 2459287.6000$^{+0.0013}_{-0.0014}$ \\
    Impact Parameter & 0.777$^{+0.021}_{-0.027}$  \\
    Radius ratio R$_p$/R$_\star$ & 0.1045$^{+0.0015}_{-0.0016}$  \\ 
    TESS limb darkening q1 & 0.480$^{+0.088}_{-0.081}$   \\
    TESS limb darkening q2 & 0.345$^{+0.083}_{-0.075}$\\ 
    NGTS limb darkening q1 & 0.389$^{+0.082}_{-0.086}$\\
    NGTS limb darkening q2 & 0.306$^{+0.084}_{-0.086}$   \\ 
    Eccentricity & 0.294$^{+0.014}_{-0.010}$  \\ 
    Argument of periastron [deg] & 3.4$^{+4.8}_{-5.8}$ \\
    Stellar density (kg.m$^{-3}$) & 1458$^{+157}_{-133}$ \\
    RV Semi-Amplitude [m.s$^{-1}$] & 46.1$^{+1.8}_{-1.8}$ \\
    \hline
    Derived parameters \\
    Planetary Radius [R$_\mathrm{J}$] & 0.928$^{+0.032}_{-0.032}$ \\ 
    Planetary Mass  [M$_\mathrm{J}$] & 0.960$^{+0.056}_{-0.056}$ \\
    Inclination [deg] & 89.483$^{+0.028}_{-0.028}$\\
    Transit duration [hrs] & 6.13$^{+0.27}_{-0.27}$\\
    Semi-major axis [au] & 0.408$^{+0.019}_{-0.019}$\\
    Equilibrium Temperature peri. [K] & 500$^{+55}_{-84}$\\
    Equilibrium Temperature apa. [K] & 274$^{+30}_{-46}$\\
    \hline
    Instrumental Parameters \\
    TESS offset [ppm] & -33$^{+18}_{-17}$\\ 
    TESS jitter [ppm] & 2.9$^{+24.8}_{-2.6}$\\
    NGTS offset [ppm] & 87$^{+20}_{-19}$ \\ 
    NGTS jitter [ppm] & 4600$^{+178}_{-180}$ \\
    EulerCam offset [ppm] & -1568$^{+209}_{-208}$ \\ 
    EulerCam jitter [ppm] & 872$^{+81}_{-73}$ \\
    CORALIE offset [m.s$^{-1}$] & 45576.5$^{+6.9}_{-7.3}$ \\
    CORALIE jitter [m.s$^{-1}$] & 0.08$^{+1.70}_{-0.07}$ \\
    FEROS offset [m.s$^{-1}$] & 45558.1$^{+5.7}_{-5.5}$\\
    FEROS jitter [m.s$^{-1}$] & 14.0$^{+6.3}_{-4.5}$\\
    HARPS offset [m.s$^{-1}$] & 45467.3$^{+1.5}_{-1.4}$ \\
    HARPS jitter [m.s$^{-1}$] & 4.5$^{+2.4}_{-4.2}$  \\
    PFS offset [m.s$^{-1}$] & -13.6$^{+1.0}_{-0.9}$ \\
    PFS jitter [m.s$^{-1}$] & 3.3$^{+1.4}_{-1.0}$  \\
    \hline                                   %inserts single line
    \\
    \end{tabular}
\end{table}

\begin{figure*}
\centering
\includegraphics[width=\hsize]{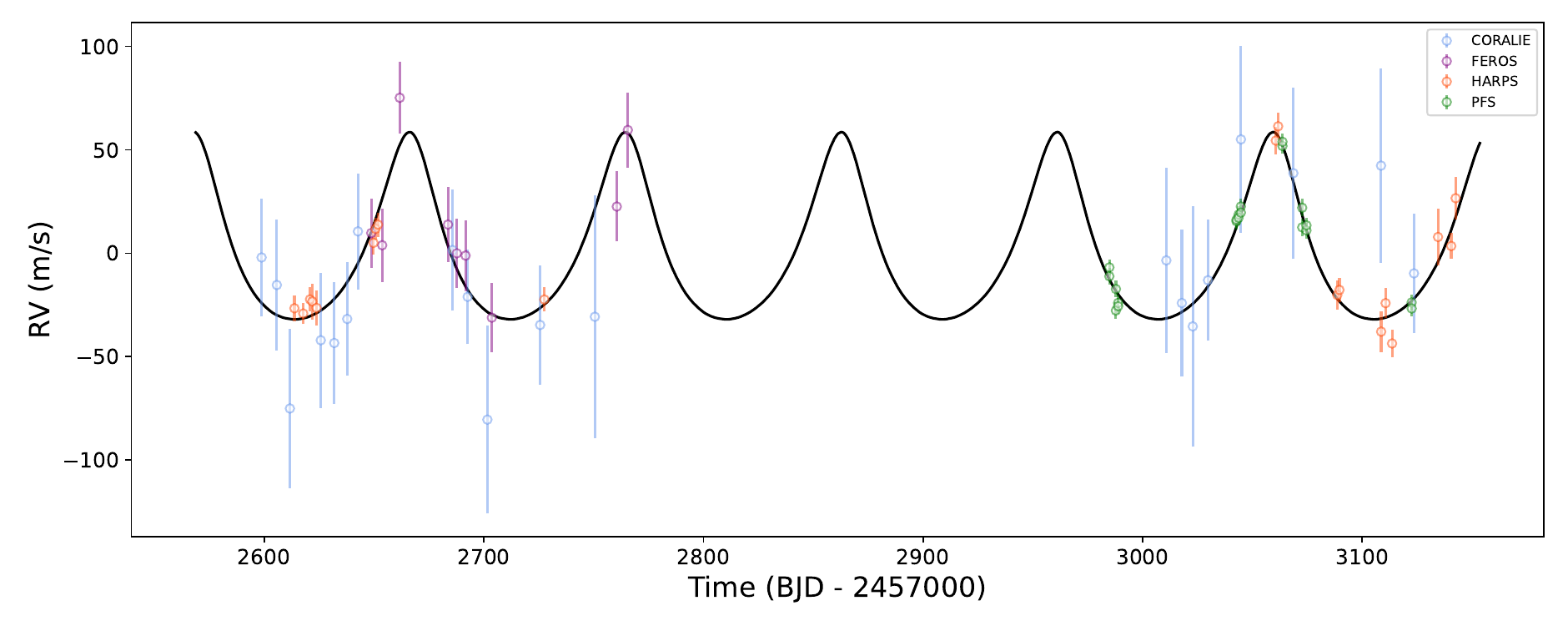}
  \caption{RV time series (in colour) with the RV model from the joint fit over-plotted in black. CORALIE RVs are shown in blue, FEROS RVs in purple, HARPS RVs in orange, and PFS RVs in green.}
     \label{fig:rv_fit_time}
\end{figure*}

%-----------------------------------------------------------------
\section{Results and discussion} \label{results}

% Overall results
The global model for NGTS-30\,b/TOI-4862\,b revealed a 98.29838 $\pm$ 0.00010 day period Jupiter-sized planet with a radius of 0.928 $\pm$ 0.032\,$R_\mathrm{J}$ and mass of 0.960 $\pm$ 0.05d\,$M_\mathrm{J}$, representing errors in radius and mass of only 3.4\% and 5.8\%, respectively. The orbit of NGTS-30\,b/TOI-4862\,b has an eccentricity of 0.294$^{+0.014}_{-0.010}$, inclination of 89.483 $\pm$ 0.028 deg and semi-major axis of 0.408$\pm$0.019 au. 

The photometric data around the two fitted transits of TOI-4862 b and the residuals after removing the joint model are shown in Figure \ref{fig:final_transits}. The higher-cadence NGTS and EulerCam data are presented in one minute bins for clarity (in green) and ten minute bins (in black), to align with the TESS data cadence. A 1\% transit depth is clearly evident in all datasets, reflecting the obtained radius ratio of R$_p$/R$_\star$ = 0.1045$\pm$0.0015. Meanwhile, the fitted RV data are illustrated in Figures \ref{fig:rv_fit} and \ref{fig:rv_fit_time}, showing the phased and time-series data, respectively. The fitted RV model shows an amplitude of K = 46.1$\pm$1.8 m s$^{-1}$, and the medium eccentricity of the planet's orbit is clearly evident from the asymmetric shape of the model. Like the photometric data, the RV data from all instruments yield flat residuals after the removal of the joint model. Furthermore, the scatter of all points are comparable to the RV error for each instrument, agreeing with the one-planet hypothesis. The complete set of fitted and derived results for this system can be found in Table \ref{results_table} and the posterior distributions for all fitted parameters can be found in Figure \ref{fig:corner_plot}.

% Bulk modelling
\subsection{Interior modelling}
NGTS-30\,b/TOI-4862\,b is a transiting warm Jupiter with precisely determined mass, radius, and age making it a good candidate to investigate its interior composition. Given its stellar insolation of 3.90 $\pm$ 0.40 $S_\oplus$  and an average equilibrium temperature of 390K, the planet is assumed not to be inflated and thus there is no degeneracy between the heavy element content and the amount of radius inflation. The planetary evolution code \texttt{completo} (Mordasini et al. 2012) is used to model the core and the envelope along with a semi-grey atmospheric model. A grid of evolution models assuming no bloating of the planet was built and coupled on this grid with a Bayesian inference model to estimate the heavy element content. The planet is modelled with a core of $\rm 10\,M_{\oplus}$ and an envelope of hydrogen and helium in which heavy elements are homogeneously mixed. The heavy elements are modelled as water following the equation of state AQUA2020 (Hadelmann et al. 2020). Hydrogen and helium are modelled with the equation of state from Chabrier and Debras (2019, CMS2019). As shown by Müller et al. 2020, the equation of state CMS2019 is known to lead to smaller radii for a given age and metallicity than the SCvH equation of state (Saumon et al. 1995). Figure~\ref{fig:Z_fit} displays the posterior distribution of the heavy element fit for TOI-4862\,b. Using this model, the planet was found to have a heavy element mass fraction of $\rm 0.23^{+0.05}_{-0.06}$, which corresponds to a mass of $85^{+15} _{-20}$ $M_{\oplus}$: 10 $M_{\oplus}$ of this mass of heavy element is the core mass, the rest is homogeneously mixed in the envelope. Assuming that the stellar metallicity scales with the iron abundance $\rm Z_{\star} = 0.0142 \times 10^{[Fe/H]}$ (Asplund et al. 2009, Miller et al. 2011), the heavy element enrichment ($\rm Z_{p} / Z_{\star} $) is $20 ^{+5} _{-6}$ and TOI-4862\,b is metal-enriched compared to its host star. This agrees well with the theoretical predictions by \citet{Thorngren2016THEPLANETS} assuming Toomre's Q parameter has a value of Q = 5 (as with their best-fit model).

\subsection{Comparison to the known population}
\begin{figure}
\centering
\includegraphics[width=\hsize]{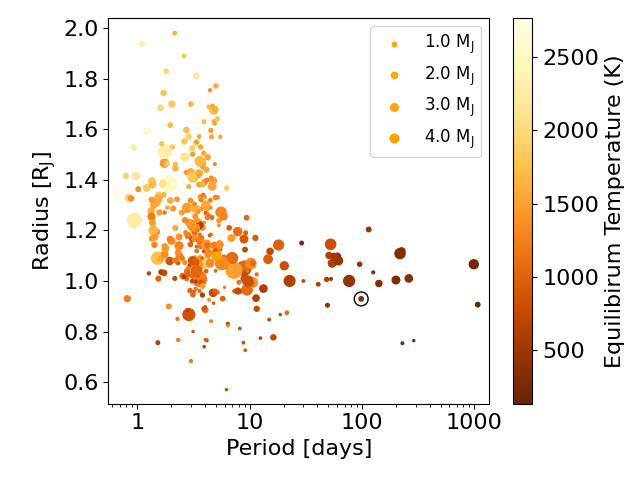}
  \caption{Period vs radius plot for all giant transiting exoplanets (0.2$R_\mathrm{J} \leq R_\mathrm{P} \leq 13 R_\mathrm{J}$)  with well-constrained masses ($\sigma_{M_P} / M_P \leq$ 25\%) and radii ($\sigma_{R_P} / R_P \leq$ 8\%). Individual points are colour-coded by the V-magnitude of the host star. TOI-4862\,b is circled in black.}
     \label{fig:per-rad}
\end{figure}

\begin{figure}
\centering
\includegraphics[width=\hsize,trim=4 4 4 4,clip]{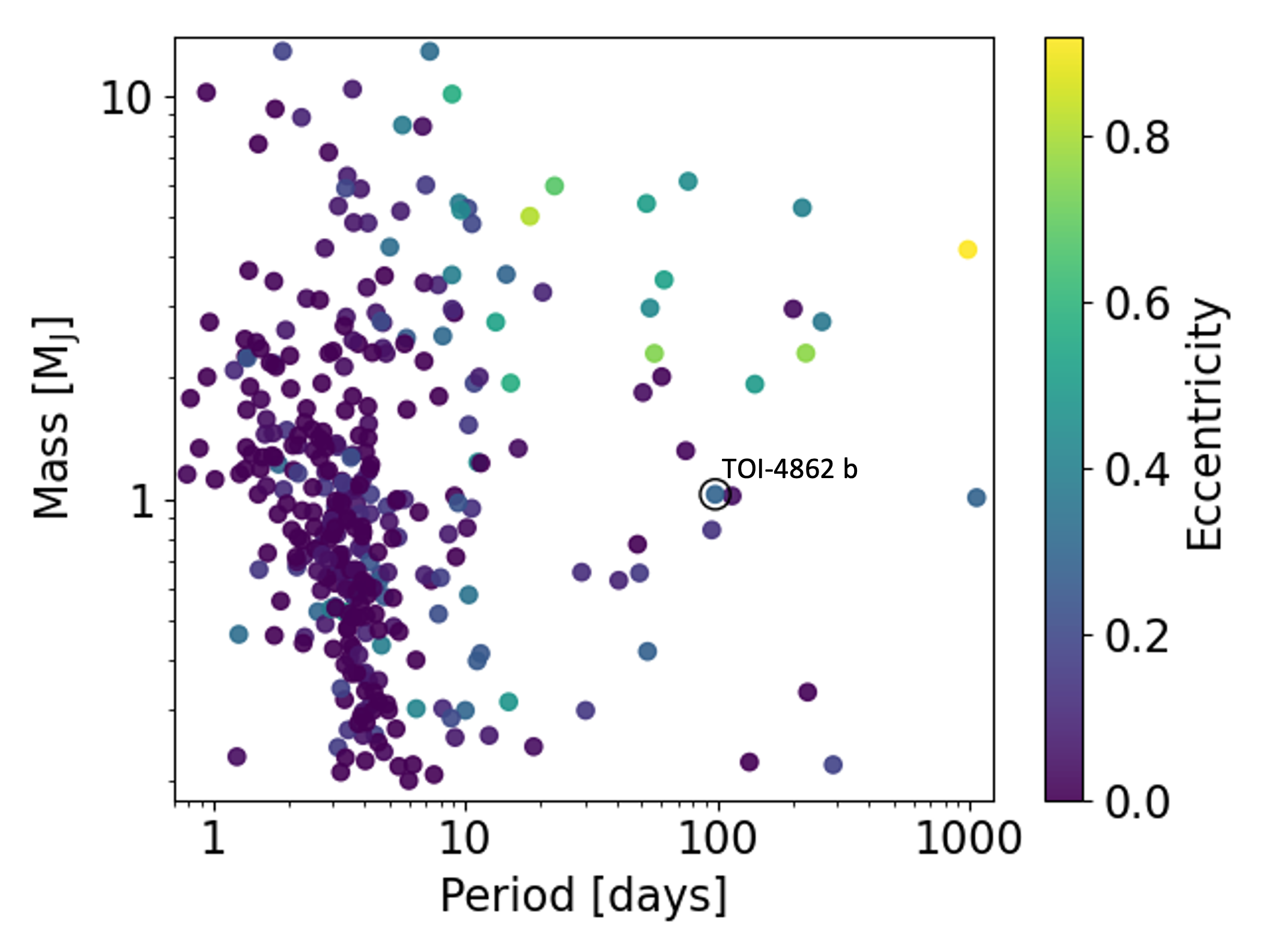}
  \caption{Period vs mass plot for all transiting giant exoplanets (0.2$R_\mathrm{J} \leq R_\mathrm{P} \leq 13 R_\mathrm{J}$) with well-constrained masses ($\sigma_{M_P} / M_P \leq$ 25\%) and radii ($\sigma_{R_P} / R_P \leq$ 8\%). Individual points are colour-coded by the planetary eccentricity, illustrating a general trend towards higher eccentricities at longer periods. TOI-4862\,b is circled in black.}
     \label{fig:per-mass}
\end{figure}

In order to place NGTS-30\,b/TOI-4862\,b in the context of other giant planet discoveries, the planet was plotted in period-radius and period-mass space in Figures \ref{fig:per-rad} and \ref{fig:per-mass}, respectively. To construct these plots, all planets from the NASA Exoplanet Archive\footnote{\url{https://exoplanetarchive.ipac.caltech.edu/}; accessed 17 November 2023} \citep{Akeson2013TheResearch} were downloaded. Then a sample was assembled by choosing those that were giant (0.2$R_\mathrm{J} \leq R_\mathrm{P} \leq 13 R_\mathrm{J}$), transiting, and had well-constrained masses ($\sigma_{M_P} / M_P \leq$ 25\%) and radii ($\sigma_{R_P} / R_P \leq$ 8\%). Plotted in colour are the V-band magnitude (Figure \ref{fig:per-rad}) and the eccentricity (Figure \ref{fig:per-mass}) of this sample. These plots illustrate that TOI-4862\,b has a small radius compared to other currently known giant exoplanets of $\sim$100 days, but a mass approximately in the middle of the population. The magnitude of its host star (Vmag = 12.5) places it close to the middle of the range, keeping it within reach of atmospheric characterisation. 

%Eccentricity
\subsection{Eccentricity}
Figure \ref{fig:per-mass} shows that NGTS-30\,b/TOI-4862\,b has a moderate eccentricity of 0.294, in line with the median eccentricity for exoplanets with periods greater than $\sim$50 days (n.b. as above, these data were drawn from the NASA Exoplanet Archive$^6$). Eccentricity is particularly interesting for warm Jupiters, as there appear to be two distinct groups of eccentricity: higher-mass (>2 M$_\mathrm{J}$) planets tend to have larger eccentricities, while lower-mass warm Jupiters generally appear to have smaller eccentricities (see Figure \ref{fig:per-mass}). A similar split in eccentricity between high mass ($\geq$ 4M$_\mathrm{J}$) and low mass (<4M$_\mathrm{J}$) giants was seen by \citet{Ribas2007TheMechanisms} in RV planet discoveries, and they suggested that this pointed towards higher mass warm Jupiters forming from fragmentation of the pre-stellar cloud (in a similar manner to binary stars) and lower-mass warm Jupiters forming via more classical core accretion mechanisms \citep[e.g.][]{Safronov1972EvolutionPlanets.,Lissauer1993PlanetFormation,Pollack1996FormationGas}. However, this discrepancy could also be caused by massive planets remaining stable on eccentric orbits while lower-mass planets become unstable \citep[e.g.][]{Chatterjee2008DynamicalScattering}. One way to further investigate the past evolutionary history of this planet would be to measure the spin-orbit angle of TOI-4862\,b, which can be achieved through observations of the Rossiter-McLaughlin effect \citep[e.g.][]{Rossiter:1924,McLaughlin:1924,Ohta2005TheSystems,Cegla2016TheSystems,Wang2021TheK2-232b,Rice:2021,Bourrier2021TheOrbits}. Such a measurement would provide a cornerstone constraint in an almost entirely unpopulated region of parameter space for warm Jupiters with 2D spin-orbit measurements -- at very long orbital periods and wide star-planet separations \citep{Rice:2022, Albrecht:2022}. Using Eq. (40) from \citet{Winn2010ExoplanetOccultations} gives an estimated value for the Rossiter-McLaughlin  effect of 24\,m/s.

% Equilibrium temperature
\subsection{Equilibrium temperature}
The current eccentricity of NGTS-30\,b/TOI-4862\,b's orbit coupled with its long period implies that interesting changes occur in its equilibrium temperature over the course of its orbit. Given its similarity to Jupiter, the equilibrium temperature of TOI-4862\,b can be estimated by assuming a Jupiter-like albedo of 0.343 and generating upper and lower bounds by assuming bond albedos of A = 0 and A = 0.686 (double that of Jupiter). Using Eq. 1 (which takes into consideration the eccentricity of the planet but assumes instant heat distribution and a vanishingly small heat capacity) the planetary equilibrium temperature was calculated to vary from 500$^{+55}_{-84}$\,K at periastron to 274$^{+30}_{-46}$\,K at apastron. Although in reality the temperature changes over the course of the planet's orbit will be tempered by the atmospheric heat capacity and redistribution of heat in the planet's atmosphere, the dramatic change in temperature over the course of this eccentric wide orbit illustrates the unusual and evolving environment faced by this exoplanet.

\begin{equation}
    T_{\mathrm{eq}} = T_{\mathrm{eff}} (1-A)^{1/4}\sqrt{\frac{R_*}{2a(1\pm e)}}
\end{equation}

\begin{figure}
\centering
\includegraphics[width=\hsize]{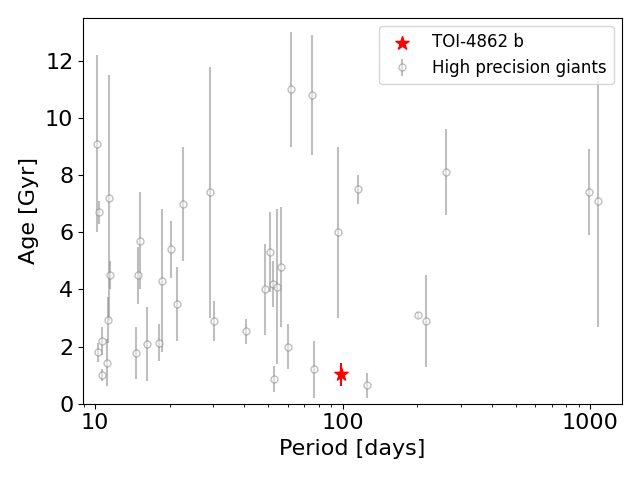}
  \caption{Period vs age plot for all giant exoplanets  (0.2$R_\mathrm{J} \leq R_\mathrm{P} \leq 13 R_\mathrm{J}$) with well-constrained masses ($\sigma_{M_P} / M_P \leq$ 25\%) and radii ($\sigma_{R_P} / R_P \leq$ 8\%), periods of >10 days, and errors in age of $\sigma_{\mathrm{Age}} / \mathrm{Age}\leq$100\%. TOI-4862\,b is shown by the red star, illustrating its position as one of the youngest warm Jupiters to date with a well-constrained age. The large age errors of most long-period giants are also evident.}
     \label{fig:per-age}
\end{figure}

\subsection{Prospects for atmospheric characterisation}

NGTS-30\,b/TOI-4862\,b joins a small but growing population of intermediate insolation giant exoplanets that are amenable to atmospheric characterisation using transmission spectroscopy. These warm and temperate gas giants are the missing link between the cold gas giants of the Solar System and the well-characterised hot Jupiter population. Because warm and temperate gas giants exist at wide enough orbits to avoid tidal circularisation, information on their migration history is retained in their eccentricity \citep{Dawson2018OriginsJupiters}. By exploring links between atmospheric composition and the mass, age, period, eccentricity, and planetary companions of warm and temperate gas giants we can hope to understand the migration and formation history of hot Jupiters.

The atmospheres of temperate gas giants like NGTS-30\,b/TOI-4862\,b ($<$500 K) are particularly interesting as they can shed light on a cooler physical and chemical atmospheric regime than has been previously studied. In particular observationally unconstrained key transitions in nitrogen chemistry \citep{Fortney2020BeyondPlanets,hu2021photochemistry,Ohno2023NitrogenSpectra} and aerosol properties \citep{gao2017sulfur,Brande2023CloudsExoplanets} are predicted to occur in this parameter space. The eccentricity of NGTS-30\,b/TOI-4862\,b means it may probe these transitions during its orbit.

With a transmission spectroscopy metric of 16, NGTS-30\,b/TOI-4862\,b is amenable to characterisation with JWST, greater than the cycle 2 temperate gas giant target PH-2b (PID 3235). This allows for the measurement of chemical composition through the C/N/O ratio, which can therefore be related to the planet's formation and migration history \citep{Mordasini2012CharacterizationFormation,piso2016role,cridland2020connecting,Ohno2023NitrogenSpectra}. The young age and well-constrained moderate eccentricity of NGTS-30\,b/TOI-4862\,b makes it a unique addition to the temperate gas giant population, which is otherwise composed of evolved planets of lower or uncertain eccentricity.

%Age
\subsection{Age}
Another key parameter for anchoring the formation and evolution of warm Jupiters is age. In this analysis NGTS-30\,b/TOI-4862\,b and its host star were found to have an age of 1.1 $\pm$ 0.4 Gyr based on the star's $v\sin i$ value, making it a fairly young, or `adolescent' system (n.b. `young' exoplanets are often deemed to be those with ages of $\leq$1Gyr - see for example \citealt{Battley2020AImages}). In order to place this discovery in the context of the wider warm Jupiter population, the aforementioned sample of high precision giant exoplanets, restricted to those giants with periods greater than 10 days and age errors of $\sigma_{\mathrm{Age}} / \mathrm{Age}\leq$100\%, was plotted in period-age parameter space in Figure \ref{fig:per-age}. What is immediately obvious is that most long-period planets have very poorly constrained ages, severely hampering modelling of their evolution through time. In order to form a reliable picture of how long-period exoplanets evolve with time, it is imperative that better age constraints are put on these planets via carefully ageing their host stars. In this regard TOI-4862 is better than most, with age errors of only 400 Myr thanks to a clear $v\sin i$ signal with which to conduct gyrochronology. The 1.1 Gyr age makes TOI-4862 b one of the youngest long-period exoplanets found to date with both a precise mass and radius, providing the community with a cornerstone system with which to probe the early evolution of warm Jupiters. Furthermore, as discussed in section \ref{method_star}, there is marginal evidence from the WASP data that the rotation period may be as low as 7.5 $\pm$ 3 days (at the shorter end of the $v \sin i$ constraints), suggesting that the planet is even younger. Placing further constraints on the age of this system and/or acquiring additional signs of youth for TOI-4862 would be very valuable for more precisely anchoring this planet within this dynamic epoch of evolution.

%Prospects for atmospheric followup 

%-----------------------------------------------------------------
\section{Conclusions} \label{conclusions}

This paper has presented the discovery and characterisation of the newly discovered long-period exoplanet NGTS-30\,b/TOI-4862\,b, an adolescent warm Jupiter on a 98.3-day orbit around the G-type star TOI-4862. This planet has a radius of 0.928 $\pm$0.032 R$_\mathrm{J}$, a mass of 0.960 $\pm$ 0.056 M$_\mathrm{J}$, and an orbital eccentricity of 0.294$^{+0.014}_{-0.010}$. NGTS-30\,b/TOI-4862\,b was discovered through a single transit from TESS before its true period was determined through a combination of photometric monitoring with NGTS and spectrographic RV measurements from CORALIE, FEROS, HARPS, and PFS, which allowed tight constraints to be placed on the planetary mass. The star's age of 1.1 Gyr makes NGTS-30\,b/TOI-4862\,b one of the youngest warm Jupiters found, adding a crucial new system with which to understand the evolution timescale of such planets. Further observations of the spin-orbit angle and the atmosphere of this planet will help further define the migration history and composition of this interesting system.

\begin{acknowledgements}
This work has been carried out within the framework of the National Centre of Competence in Research PlanetS supported by the Swiss National Science Foundation under grants 51NF40\_182901 and 51NF40\_205606. The authors acknowledge the financial support of the SNSF. 
Based on data collected under the NGTS project at the ESO La Silla Paranal Observatory. The NGTS facility is operated by the consortium institutes with support from the UK Science and Technology Facilities Council (STFC) under projects ST/M001962/1, ST/S002642/1 and ST/W003163/1.
This work has made use of data from the European Space Agency (ESA) mission {\it Gaia} (\url{https://www.cosmos.esa.int/gaia}), processed by the {\it Gaia}
Data Processing and Analysis Consortium (DPAC,
\url{https://www.cosmos.esa.int/web/gaia/dpac/consortium}). Funding for the DPAC has been provided by national institutions, in particular the institutions participating in the {\it Gaia} Multilateral Agreement. Funding for the TESS mission is provided by NASA’s Science Mission Directorate. We acknowledge the use of public TESS data from pipelines at the TESS Science Office and at the TESS Science Processing Operations Center. This paper includes data collected by the TESS mission that are publicly available from the Mikulski Archive for Space Telescopes (MAST). Resources supporting this work were provided by the NASA High-End Computing (HEC) Program through the NASA Advanced Supercomputing (NAS) Division at Ames Research Center for the production of the SPOC data products. This research has made use of the Exoplanet Follow-up Observation Program website, which is operated by the California Institute of Technology, under contract with the National Aeronautics and Space Administration under the Exoplanet Exploration Program. This research has made use of the NASA Exoplanet Archive, which is operated by the California Institute of Technology, under contract with the National Aeronautics and Space Administration under the Exoplanet Exploration Program.
JSJ greatfully acknowledges support by FONDECYT grant 1201371 and from the ANID BASAL project FB210003.
ML acknowledges support of the Swiss National Science Foundation under grant number PCEFP2\_194576.
KAC acknowledges support from the TESS mission via subaward s3449 from MIT.
This research was funded in part by the UKRI, (Grants ST/X001121/1, EP/X027562/1)
HC acknowledges the support of the Swiss National Science Foundation under grant number PCEFP2\_194576.
AO is funded by an STFC studentship. 
MR acknowledges support from Heising-Simons Foundation Grant \#2023-4478.
E.G. gratefully acknowledges support from the UK Science and Technology Facilities Council (STFC; project reference ST/W001047/1).
T.D. acknowledges support by the McDonnell Center for the Space Sciences at Washington University in St. Louis.
MK acknowledges support from the MIT Kavli Institute as a Juan Carlos Torres Fellow.
\end{acknowledgements}

\bibliographystyle{aa} % style aa.bst
\bibliography{references.bib,kgs_references.bib} % your references Yourfile.bib
\appendix

\section{Joint modelling priors}

%\begin{center}
\begin{table*}[htb!]
\centering                          % used for centering table
\caption{Priors for the joint modelling of photometric and RV data.}             % title of Table
\label{joint_priors_table}      % is used to refer this table in the text
    \begin{tabular}{l l c}        % centered columns (3 columns)
    \hline\hline                 % inserts double horizontal lines
    Parameter  & Distribution & Value \\    % table heading 
    \hline                        % inserts single horizontal line
    Orbital Period [days] & Normal & (98.3, 1.5) \\
    Mid-transit time $T_0$ [days] & Normal & (2459287.603184,0.2)  \\
    Impact Parameter & Uniform & (0, 1) \\
    Radius ratio R$_p$/R$_\star$ & Uniform & (0,1) \\ 
    TESS limb darkening q1 & Normal & (0.363879, 5 x 0.017721)  \\
    TESS limb darkening q2 & Normal  &  (0.291963, 5 x 0.017069) \\ 
    NGTS limb darkening q1 & Normal & (0.417738, 5 x 0.017562)  \\
    NGTS limb darkening q2 & Normal  &  (0.309776, 5 x 0.017689) \\ 
    EulerCam limb darkening q1 & Normal & (0.420494, 5 x 0.017562)  \\
    EulerCam limb darkening q2 & Normal  &      (0.310417, 5 x 0.017689) \\ 
    Eccentricity & Uniform & (0.0, 0.8)\\
    Argument of periastron [deg] & Uniform & (0, 360)\\
    Stellar density (kg.m$^{-3}$) & Normal & (1741.325, 200)\\
    TESS offset [ppm] & Normal & (0.0,100000)\\ 
    TESS jitter [ppm] & LogUniform & (0.1, 1000)\\
    NGTS offset [ppm] & Normal & (0.,100000)\\ 
    NGTS jitter [ppm] & LogUniform & (0.1, 10000)\\
    EulerCam offset [ppm] & Normal & (0.,100000)\\ 
    EulerCam jitter [ppm] & LogUniform & (0.1, 10000)\\
    RV Semi-Amplitude [m.s$^{-1}$] & Uniform & (0,100)\\
    Spectrograph Jitters [m.s$^{-1}$] & LogUniform & (0.001,100) \\
    CORALIE offset [m.s$^{-1}$] & Uniform & [40000,50000]\\
    FEROS offset [m.s$^{-1}$] & Uniform & [40000,50000]\\
    HARPS offset [m.s$^{-1}$] & Uniform & [40000,50000]\\
    PFS offset [m.s$^{-1}$] & Uniform & [-100,100]\\
    \hline                                   %inserts single line
    \\
    \end{tabular}
    \vspace{1ex}
\end{table*}
%\end{center}

\section{Joint modelling posterior distributions}

\begin{figure*}[htb!]
\resizebox{\hsize}{!}{\includegraphics[width=6in,trim=4 4 4 4,clip]{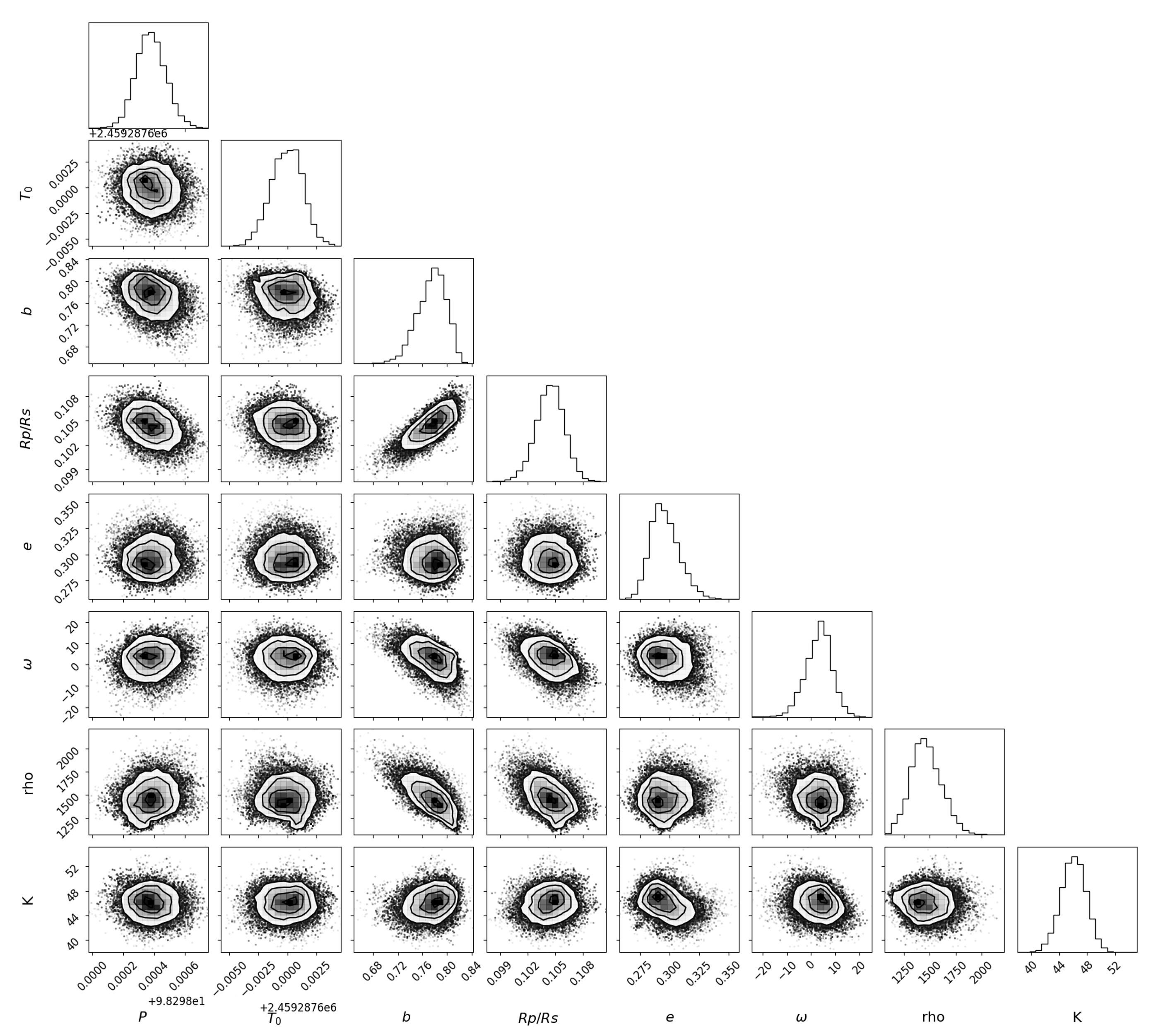}}
  \caption{Posterior distribution for fitted parameters from the joint fit of NGTS-30\,b/TOI-4862\,b.}
     \label{fig:corner_plot}
\end{figure*}

\section{Interior modelling posterior distributions}

\begin{figure*}
%\centering
\includegraphics[width=\hsize]{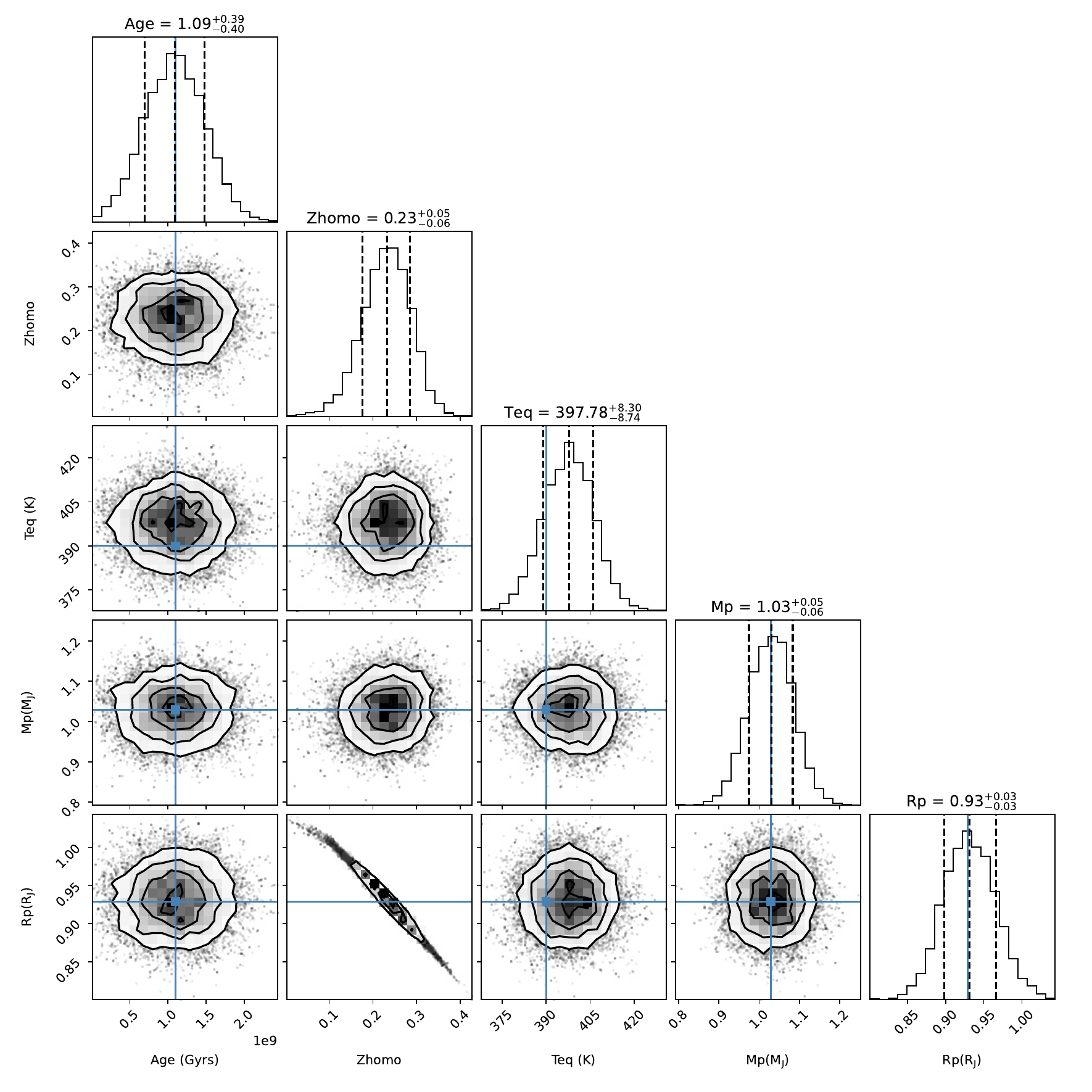}
  \caption{Posterior probability distribution for the interior modelling of TOI-4862\,b. The age is in years, Zhomo (the homogeneous metallicity) corresponds to the fraction of heavy elements in the atmosphere, and the planetary mass and radius are in Jupiter units. Blue lines mark the mean values for each parameter.}
    \label{fig:Z_fit}
\end{figure*}

\end{document}